\documentclass{article}
\usepackage{textcomp}
\usepackage[utf8]{inputenc}
\usepackage{xcolor}
\usepackage{float}
\usepackage{amsmath}
\usepackage{amsthm}
\usepackage{graphicx}
\usepackage[bookmarks=false,
 breaklinks=false,pdfborder={0 0 1},backref=section,colorlinks=false]
 {hyperref}
\hypersetup{
 hidelinks}
\usepackage[switch]{lineno}

\makeatletter

\newcommand*\LyXZeroWidthSpace{\hspace{0pt}}
\providecommand{\tabularnewline}{\\}
\floatstyle{ruled}
\newfloat{algorithm}{tbp}{loa}
\providecommand{\algorithmname}{Algorithm}
\floatname{algorithm}{\protect\algorithmname}

\usepackage{ijcai26}

\usepackage{times}
\usepackage{soul}
\usepackage{url}
\usepackage[small]{caption}
\usepackage{graphicx}
\usepackage{amsthm}

\usepackage{comment}

\usepackage{textcomp}
\usepackage[utf8]{inputenc}
\usepackage{amsmath}
\usepackage{amsthm}
\usepackage{amssymb}

\makeatletter

\providecommand{\tabularnewline}{\\}

\theoremstyle{plain}

\theoremstyle{remark}

\makeatother

\usepackage{babel}
\providecommand{\remarkname}{Remark}
\providecommand{\theoremname}{Theorem}

\title{Study and Improvement of Search Algorithms in Multi-Player Perfect-Information Games}

\author{
Quentin Cohen-Solal$^1$\\
\affiliations
$^1$LAMSADE, Université Paris-Dauphine, PSL, CNRS, Paris, France\\
\emails
quentin.cohen-solal@dauphine.psl.eu 
}

\makeatother

\begin{document}
\maketitle
\begin{abstract}
In this article, we generalize Unbounded Minimax, the state-of-the-art
search algorithm for zero sums two-player games with perfect information
to the framework of multiplayer games with perfect information. We
experimentally show that this generalized algorithm also achieves
better performance than the main multiplayer search algorithms.
\end{abstract}


\global\long\def\minibaln{\mathrm{Minibal_{n}}}%
\global\long\def\minibalp{\mathrm{Minibal_{+}}}%
\global\long\def\minibalc{\mathrm{Minibal_{c}}}%
\global\long\def\indicatrice#1#2{\boldsymbol{1}_{#1}\left(#2\right)}%

\global\long\def\et{\ \wedge\ }%
\global\long\def\terminal#1{\mathrm{terminal}\left(#1\right)}%
\global\long\def\joueur{\mathrm{j}}%
\global\long\def\joueurUn{\mathrm{1}}%
\global\long\def\joueurDeux{\mathrm{2}}%
\global\long\def\fbin{\mathrm{f_{b}}}%
\global\long\def\Actions{\mathcal{A}}%
\global\long\def\actions#1{\mathrm{actions}\left(#1\right)}%
\global\long\def\etats{\mathcal{S}}%
\global\long\def\random{\mathrm{random}\left(\right)}%
\global\long\def\premier#1{\mathrm{first_{\text{–}}player}\left(#1\right)}%
\global\long\def\balanced#1{\mathrm{balanced_{\text{–}}player}\left(#1\right)}%
\global\long\def\rootplayer#1{\mathrm{root_{\text{–}}player}\left(#1\right)}%
\global\long\def\terminalrandom{\mathrm{t_{r}}}%
\global\long\def\hfb{\mathrm{b_{t}}}%
\global\long\def\hfp{\mathrm{p_{t}}}%
\global\long\def\hadapt{f_{\theta}}%
\global\long\def\hadaptnum#1{f_{\theta_{#1}}}%
\global\long\def\itreeset{\mathbf{T_{i}}}%
\global\long\def\treeset{\mathbf{T}}%
\global\long\def\rootset{\mathbf{R}}%
\global\long\def\hterminal{f_{\mathrm{t}}}%
\global\long\def\rnap{\mathrm{p_{rna}}}%
\global\long\def\rnab{\mathrm{b_{rna}}}%
\global\long\def\ubfm{\mathrm{UBFM}}%
\global\long\def\ubfmt{\ubfm_{\mathrm{s}}}%
\global\long\def\argmax{\operatorname{\mathrm{arg\,max}}}%
\global\long\def\argmin{\operatorname{\mathrm{arg\,min}}}%
\global\long\def\liste#1#2{\left\{  #1\,|\,#2\right\}  }%
\global\long\def\fterminal{\hterminal}%
\global\long\def\fadapt{\hadapt}%
\global\long\def\id{\mathrm{ID}}%
\global\long\def\minimum{\operatorname*{\mathrm{min}}}%
\global\long\def\Argmax#1{\operatorname{\mathrm{arg\,max}}_{#1}}%
\global\long\def\Argmin#1{\operatorname{\mathrm{arg\,min}}_{#1}}%
\global\long\def\tmax{t_{\mathrm{max}}}%
\global\long\def\negation{\mathop{\neg}}%
\global\long\def\Max#1#2{{\displaystyle \mathop{\mathrm{max}}_{#2}\left(#1\right)}}%
\global\long\def\Min#1#2{{\displaystyle \mathop{\mathrm{min}}_{#2}\left(#1\right)}}%
\global\long\def\rollout#1{\mathrm{rollout}\left(#1\right)}%
\global\long\def\et{\ \wedge\ }%
\global\long\def\ou{\ \vee\ }%
\global\long\def\alphabeta{\alpha\beta}%
\global\long\def\pvs#1{\mathrm{PVS}^{#1}}%
\global\long\def\mtdf#1{\mathrm{MTD(f)}^{#1}}%
\global\long\def\mctsh#1{\mathrm{MCTS}_{\mathrm{h}}^{#1}}%
\global\long\def\mcts#1{\mathrm{MCTS}^{#1}}%
\global\long\def\mctsip#1#2{\mathrm{MCTS}_{\gamma=#2}^{#1}}%
\global\long\def\mctsipm#1{\mathrm{MCTS}_{\alpha\beta}^{#1}}%
\global\long\def\mctsic#1{\mathrm{MCTS}_{\mathrm{max}}^{#1}}%
\global\long\def\joueur{\mathrm{j}}%
\global\long\def\joueurUn{\mathrm{1}}%
\global\long\def\joueurDeux{\mathrm{2}}%
\global\long\def\fbin{\mathrm{f_{b}}}%
\global\long\def\Actions#1{\mathrm{actions}\left(#1\right)}%
\global\long\def\zip{\mathcal{\mathrm{zip}}}%
\global\long\def\etats{\mathcal{S}}%
\global\long\def\mv#1#2{v_{#1,#2}}%
\global\long\def\smv#1{v_{#1}}%
\global\long\def\bv#1#2{v_{#1,#2}}%
\global\long\def\select#1#2{n_{#1,#2}}%
\global\long\def\sselect#1{n_{#1}}%
\global\long\def\sign#1#2{\overline{#1}^{#2}}%
\global\long\def\vsign#1#2{\overline{\mv{#1}{#2}}}%
\global\long\def\bvsign#1#2{\ddddot{\bv{#1}{#2}}}%
\global\long\def\vinf#1#2{v_{#1,#2}^{-}}%
\global\long\def\vinfsign#1#2{\overline{\vinf{#1}{#2}}}%
\global\long\def\vsup#1#2{v_{#1,#2}^{+}}%
\global\long\def\vsupsign#1#2{\overline{\vsup{#1}{#2}}}%
\global\long\def\vsinf#1{v_{#1}^{-}}%
\global\long\def\vssup#1{v_{#1}^{+}}%
\global\long\def\vresol#1#2{c_{#1,#2}}%
\global\long\def\svresol#1{c_{#1}}%
\global\long\def\resol#1#2{r_{#1,#2}}%
\global\long\def\sresol#1{r_{#1}}%
\global\long\def\vresolsign#1#2{\overline{\vresol{#1}{#2}}}%
\global\long\def\svresolsign#1{\overline{\svresol{#1}}}%
\global\long\def\mtdflower{f_{-}}%
\global\long\def\mtdfupper{f_{+}}%
\global\long\def\terminalrandom{\mathrm{t_{r}}}%
\global\long\def\hfb{\mathrm{b_{t}}}%
\global\long\def\hfp{\mathrm{p_{t}}}%
\global\long\def\hadapt{f_{\theta}}%
\global\long\def\hadaptnum#1{f_{\theta_{#1}}}%
\global\long\def\itreeset{\mathbf{T_{i}}}%
\global\long\def\treeset{\mathbf{T}}%
\global\long\def\rootset{\mathbf{R}}%
\global\long\def\hterminal{f_{\mathrm{t}}}%
\global\long\def\rnap{\mathrm{p_{rna}}}%
\global\long\def\rnab{\mathrm{b_{rna}}}%
\global\long\def\ubfm{\mathrm{UBFM}}%
\global\long\def\ubfmt{\ubfm_{\mathrm{s}}}%
\global\long\def\ubfmref{\mathrm{UBFM}_{\mathrm{ref}}}%
\global\long\def\ubfmc{\mathrm{UBFM}_{\mathrm{c}}}%
\global\long\def\ubfms{\mathrm{UBFM}_{\mathrm{s}}}%
\global\long\def\ubfmcs{\mathrm{UBFM}_{\mathrm{c-s}}}%
\global\long\def\liste#1#2{\left\{  #1\,|\,#2\right\}  }%
\global\long\def\fterminal{\hterminal}%
\global\long\def\fadapt{\hadapt}%
\global\long\def\id{\mathrm{ID}}%
\global\long\def\pdiscret{\delta}%
\global\long\def\ended#1{\mathrm{ended}\left(#1\right)}%
\global\long\def\corum{\mathrm{CORUM}}%
\global\long\def\corsum{\mathrm{CORSUM}}%
\global\long\def\actions#1{\mathrm{actions}\left(#1\right)}%

\global\long\def\terminal#1{\mathrm{ended}\left(#1\right)}%

\global\long\def\player#1{\mathrm{player}\left(#1\right)}%

\global\long\def\time{\mathrm{time}\left(\right)}%

\global\long\def\premier#1{\mathrm{first\ player}\left(#1\right)}%
\global\long\def\orsum{\mathrm{ORSUM}}%
\global\long\def\maxn{\mathrm{Max}^{n}}%
\global\long\def\brs{\mathrm{BRS}}%
\global\long\def\brsp{\mathrm{BRS}^{+}}%

\global\long\def\time{\mathrm{time}\left(\right)}%

\section{Introduction}

Adversarial search in perfect-information games is a central area
of artificial intelligence, driving significant theoretical and practical
advances. For two-player zero-sum games, search algorithms such as
Minimax and its numerous extensions have established a robust conceptual
framework, enabling effective methods over large combinatorial state
spaces. Among these extensions, Unbounded Minimax~\cite{korf1996best,cohen2020learning}
has recently emerged as a state-of-the-art algorithm~\cite{cohen2025study}. 

Unbounded Minimax explores the most promising action sequences in
a best-first manner, without being constrained by a depth limit as
in standard minimax searches. It forms, with its safe decision variant,
the core search component of the Athénan architecture~\cite{korf1996best,cohen2020learning},
the leading software in international game competitions. Over six
years, Athénan won 57 gold medals at the Computer Olympiad~\cite{cohen2021descent,cohen2023athenan,cohen2025athenan},
tripled the record for gold medals in a single year~\cite{cohen2023athenan},
and held the title in 20 games. Its main competitor, AlphaZero~\cite{silver2018general},
relies on Monte Carlo Tree Search~\cite{Coulom06}. Experimental
evaluations demonstrate that Athénan outperforms AlphaZero by up to
30 times, at least in the studied context (reasonable computing power)~\cite{cohen2023minimax}.

Extending these successes to the multi-player setting, however, remains
an open challenge. Games with more than two players, even under perfect
information, introduce fundamental structural difficulties: the absence
of simple duality, the presence of multiple objectives, non-transitive
preferences, impact of tie-breaking strategies, explosion of possible
equilibria, and several mathematical properties are lossed such as
the minimax theorem. These characteristics make most direct generalizations
of algorithms designed for two-player zero-sum games inadequate. In
particular, in multi-player settings, it appears infeasible to anticipate
many turns in advance without introducing simplifications that generate
strong biases. In addition, extending an algorithm that relies on
scalar minimax values to a vector-valued setting is not straightforward,
as dominance relations are only partial and may be incomparable.

Furthermore, many multiplayer algorithms have been proposed, but their
relative performance is an open problem. It is unknown whether a better
multiplayer algorithm exists, at least with regard to average performance,
and if not, under what conditions one algorithm outperforms another.

In this paper, we propose a novel search algorithm for multi-player
perfect-information games that generalizes Unbounded Minimax. Our
approach preserves the key principles that make Unbounded Minimax
effective: unbounded search, incremental state resolution, and the
prioritization of critical action sequences, while adapting them to
a multi-objective, non-strictly antagonistic setting. This allows
us to mitigate the combinatorial explosion, which is even more severe
in multi-player games, and thereby enables longer-term planning in
this generalized context. In addition, we perform an experimental
evaluation across several benchmark multi-player games, demonstrating
that the proposed algorithm surpasses the average performance of other
multiplayer algorithms.

These results open the way to a unified conceptual treatment of adversarial
search in perfect-information games, covering both two-player and
multi-player cases while maintaining competitive large-scale performance.

The remainder of this paper is organized as follows. In Section~\ref{sec:Related-Work},
we review related work, covering Unbounded Minimax and classical multi-player
game algorithms. Section~\ref{subsec:Unbounded_maxn} formalizes
our new search algorithms: Unbounded $\maxn$ and Unbounded $\maxn$
with safe decision. In Section~\ref{sec:Experiments}, we describe
our experimental protocol, including the evaluation setup, technical
details, and the set of games considered. We then present its results
in Section~\ref{subsec:Results-of-the}. Finally, Section~\ref{sec:Conclusion}
concludes the paper and outlines directions for future work.

\section{Related Work}

\label{sec:Related-Work}
\begin{table*}[t]
\begin{centering}
{\footnotesize{}%
\begin{tabular}{|c|c|}
\hline 
{\footnotesize Symbols} & {\footnotesize Definition}\tabularnewline
\hline 
\hline 
$s$ & {\footnotesize a game state}\tabularnewline
\hline 
$a$ & {\footnotesize an action of the game}\tabularnewline
\hline 
{\footnotesize$\actions s$} & {\footnotesize action set of the state $s$ for the current player}\tabularnewline
\hline 
{\footnotesize$\premier s$} & {\footnotesize true if the current player of the state $s$ is the
first player}\tabularnewline
\hline 
{\footnotesize$\mathrm{player}\left(s\right)$} & {\footnotesize number of the player that must play in $s$}\tabularnewline
\hline 
{\footnotesize$\terminal s$} & {\footnotesize true if $s$ is an end-game state}\tabularnewline
\hline 
{\footnotesize$a(s)$} & {\footnotesize state obtained after playing the action $a$ in the
state $s$}\tabularnewline
\hline 
{\footnotesize$\time$} & {\footnotesize current time in seconds}\tabularnewline
\hline 
{\footnotesize$S$} & {\footnotesize keys of the transposition table $T$}\tabularnewline
\hline 
{\footnotesize$T$} & {\footnotesize transposition table: $T=\liste{\left(c_{s,a},\mv sa,\select sa,r_{s,a}\right)}{s\in S\et a\in\mathrm{actions}(s)}$}\tabularnewline
\hline 
{\footnotesize$\tmax$} & {\footnotesize search time per action}\tabularnewline
\hline 
{\footnotesize$t$} & {\footnotesize time elapsed since the start of the search}\tabularnewline
\hline 
{\footnotesize$\select sa$} & {\footnotesize number of times the action $a$ is selected in state
$s$}\tabularnewline
\hline 
{\footnotesize$\mv sa$} & {\footnotesize partial minimax value obtained after playing action
$a$ in state $s$}\tabularnewline
\hline 
$\vresol sa$ & {\footnotesize completion value obtained after playing action $a$
in state $s$}\tabularnewline
\hline 
$\resol sa$ & {\footnotesize resolution value obtained after playing action $a$
in state $s$}\tabularnewline
\hline 
{\footnotesize$\hadapt(s)$} & {\footnotesize heuristic evaluation function (of non-terminal game
tree leaves)}\tabularnewline
\hline 
{\footnotesize$\hterminal(s)$} & {\footnotesize evaluation of terminal states, e.g. game gain or score}\tabularnewline
\hline 
{\footnotesize$\hfb(s)$} & {\footnotesize Two-players gain function: $1$ if the first player
wins, $0$ in case of a draw, -1 if the first player loses}\tabularnewline
\hline 
{\footnotesize$\hfb(s)$} & {\footnotesize Multi-players gain function: $\hfb(s)_{p}=1$ if player
$p$ wins, $\hfb(s)_{p}=-1$ if player $p$ loses, $\hfb(s)_{p}=0$
otherwise.}\tabularnewline
\hline 
\end{tabular}}{\footnotesize\par}
\par\end{centering}
\caption{Index of symbols}\label{tab:Index-of-symbols}
\end{table*}

\subsection{Best Action Searching in Game Trees}

The vast majority of game-search algorithms rely on representing a
game as a tree, where nodes correspond to game states and edges correspond
to actions that allow transitions from one state to another. The terminal
nodes represent end-game states and are assigned multi-dimensional
values according to the final game outcome. Typically $v\in\left\{ -1,0,1\right\} ^{n}$
where $n$ is the number of players and $v_{j}=1$ means player $j$
wins, $v_{j}=0$ means draw, and $v_{j}=-1$ means player $j$ loses.
More generally, $v\in\mathbb{R}^{n}$ and $v_{j}$ is the score of
player $j$. This valuation can be propagated to the other nodes under
the assumption that all players play optimally. However, this theoretical
valuation is generally intractable in practice, as game trees represent
enormous state spaces. Consequently, algorithms must estimate these
multi-dimensional values or the corresponding strategies to make practical
multi-player game-playing decisions.

\subsection{Base Multi-player Searches: $\protect\maxn$, Paranoid, $k$-best
pruning}

\label{subsec:Classical-adversarial-search}

The $\maxn$ algorithm~\cite{luckhart1986algorithmic} is the canonical
approach for deterministic multi-player games: it constructs a game
tree to a fixed depth, evaluates the leaf nodes with a heuristic evaluation
function, and propagates these values back up the tree under the assumption
that all players independently maximize their own payoff. $\maxn$
remains the simplest and most fundamental framework for reasoning
about multi-player adversarial behavior. In $\maxn$, the value of
a state where the player $j$ must play is the value $v\left(s\right)$
of its child-state $s$ whose component $v\left(s\right)_{j}$ is
maximum.

Unfortunately, this multiplayer generalization of the minimax algorithm
causes it to lose many of its properties. In particular, several standard
minimax improvements are no longer applicable in this context, such
as alpha-beta pruning~\cite{knuth1975analysis}, which significantly
reduces combinatorial explosion. 

An alternative approach, known as Paranoid search~\cite{sturtevant2000pruning},
reduces the multi-player setting to a two-player zero-sum game by
assuming that all opponents form a coalition against the player which
must play (the root player). Although this pessimistic assumption
often enables deeper searches, particularly through the application
of alpha--beta pruning, it introduces strong biases and can lead
to overly conservative play.

Another technique used to circumvent combinatorial explosion is $k$-best
pruning~\cite{baier2018mcts}. This involves keeping only the $k$
best actions a priori, which artificially reduces the branching factor
to $k$ and thus allows for longer-term planning. This pruning can
remove optimal strategies; it is an inexact pruning, unlike the alpha-beta
pruning of the two-player framework. Another drawback of this technique
is that it is parameterized by a constant $k$ which needs to be tuned.

\subsection{Best Reply Search: $\protect\brs$ and $\protect\brsp$}

Best-Reply Search ($\brs$) was proposed as a novel compromise between
$\maxn$and Paranoid search.

$\brs$ restricts opponents’ behavior by allowing only the opponent
with the strongest counter-move to act between two consecutive turns
of the root player, while the other players do not play. This significantly
reduces the branching factor and enables deeper lookahead for the
root player. Empirical evaluations demonstrated that $\brs$ can outperform
$\maxn$ and often surpass Paranoid search in games such as Chinese
Checkers, Focus, and Rolit.

However, the original formulation of $\brs$ may generate invalid
game states by violating the proper turn order among opponents. This
consequently requires modifying the game implementation so that these
invalid states can be played, which is not always technically possible
and is costly in terms of human resources. This limitation was addressed
by $\brsp$, which refines the original algorithm to preserve turn
consistency during the search. $\brsp$ relies on move ordering to
select representative opponent moves that are not explicitly searched,
thereby maintaining valid game trajectories while retaining the computational
advantages of $\brs$. With $\brsp$, between two consecutive turns
of the root player, one opponent acts normally, while the others play
their representative moves. Experimental results in Four-Player Chess
show that $\brsp$ significantly improves upon $\brs$, achieving
higher win rates against $\maxn$, Paranoid, and $\brs$ itself, at
least in some contexts.

\subsection{Monte-Carlo Tree Search: $\protect\mcts{}$ and $\protect\mctsh{}$}

\label{subsec:Monte-Carlo-Tree-Search}

Monte Carlo Tree Search (MCTS) \cite{browne2012survey,Coulom06} reframed
game search for domains where hand-crafted evaluation functions are
inadequate. Instead of evaluating every leaf node with a heuristic,
MCTS performs randomized simulations (playouts) and uses their outcomes
to guide a best-first expansion of the search tree through four iterative
steps: selection, expansion, simulation, and backpropagation. MCTS
has achieved remarkable success across a wide range of board and real-time
games. $\mcts{}$ is parameterized by a constant $C$ which needs
to be tuned (theoretical value $\sqrt{2}$).

It has been extended to the multiplayer framework, in the same way
that minimax has been generalized to $\maxn$~\cite{sturtevant2008analysis}.

It has also been modified in various ways to incorporate knowledge
to improve its performance by biasing its selections and/or simulations~\cite{browne2012survey}.
The simplest method is to replace state evaluation by playouts with
heuristic evaluation~\cite{ramanujan2011trade}. We denote this variant
by $\mctsh{}$.

\subsection{Unbounded Minimax and Safe Decision }

\label{subsec:Athenan}

\subsubsection{Unbounded Minimax}

Unbounded (Best-First) Minimax \cite{cohen2025study,korf1996best,cohen2025little,cohen2020learning,cohen2021completeness}
is a variant of the classical Minimax algorithm for two-player, zero-sum,
perfect-information games. Standard Minimax explores all possible
game evolutions up to a fixed depth $d$, evaluating states under
the assumption that the first player maximizes its payoff while the
second player minimizes it. However, for large $d$, exhaustive exploration
becomes impractical, as the number of states grows exponentially with
depth.

Unbounded Minimax addresses this limitation by replacing uniform depth-limited
exploration with a best-first, non-uniform search strategy. Rather
than expanding all states, it selectively explores those estimated
most promising, enabling very deep local lookahead. More specifically,
it determines the best current sequence of explored actions and extends
it by one action and iteratively restarts this process. Importantly,
Unbounded Minimax is not parameterized by a predefined search depth:
it incrementally extends the most relevant action sequences by one
action and can be interrupted at any time, yielding a valid current
solution.

\subsubsection{Safe Unbounded Minimax}

Unbounded Minimax was subsequently enhanced with the safe decision
technique~\cite{cohen2025study,cohen2020learning}. This approach
combines the standard Unbounded Minimax search procedure with the
\textsl{safe decision strategy}. The safe decision is a variant for
the final action-selection rule. Thus, in this variant, only the criterion
used to choose the action after the search is modified. In the base
Unbounded Minimax, the selected action is the one that leads to the
child state with the best value. Under the safe decision strategy,
however, the action played is the safest one, defined as the action
most frequently selected during the search. In other words, the chosen
action is the one that maximizes the number of times it has been selected
from the root. Because the search is conducted in best-first order,
the most frequently selected action is also the one that was most
often evaluated as best throughout the search. This leads to robustness
against evaluation errors. Moreover, it is the action for which the
algorithm has the greatest visibility over possible future developments,
as its corresponding subtree has been explored most extensively. 

\subsubsection{Unbounded Minimax with Completion}

Unbounded Minimax has also been improved by the completion technique~\cite{cohen2021completeness,cohen2020learning}.
First, it avoids exploring actions that lead to solved states (i.e.,
states whose exact value is known; in other words, further exploration
of their subtrees can never change their value), which prevents the
algorithm from getting stuck in suboptimal fixed points. Second, it
always selects after the search an action leading to a solved winning
state and never selects, when avoidable, an action leading to a solved
losing state.

Unbounded Minimax with completition (without and with Safe Decision)
is described in Algorithm~\ref{alg:ubfms}

\begin{algorithm}
\begin{centering}
\includegraphics[scale=0.5]{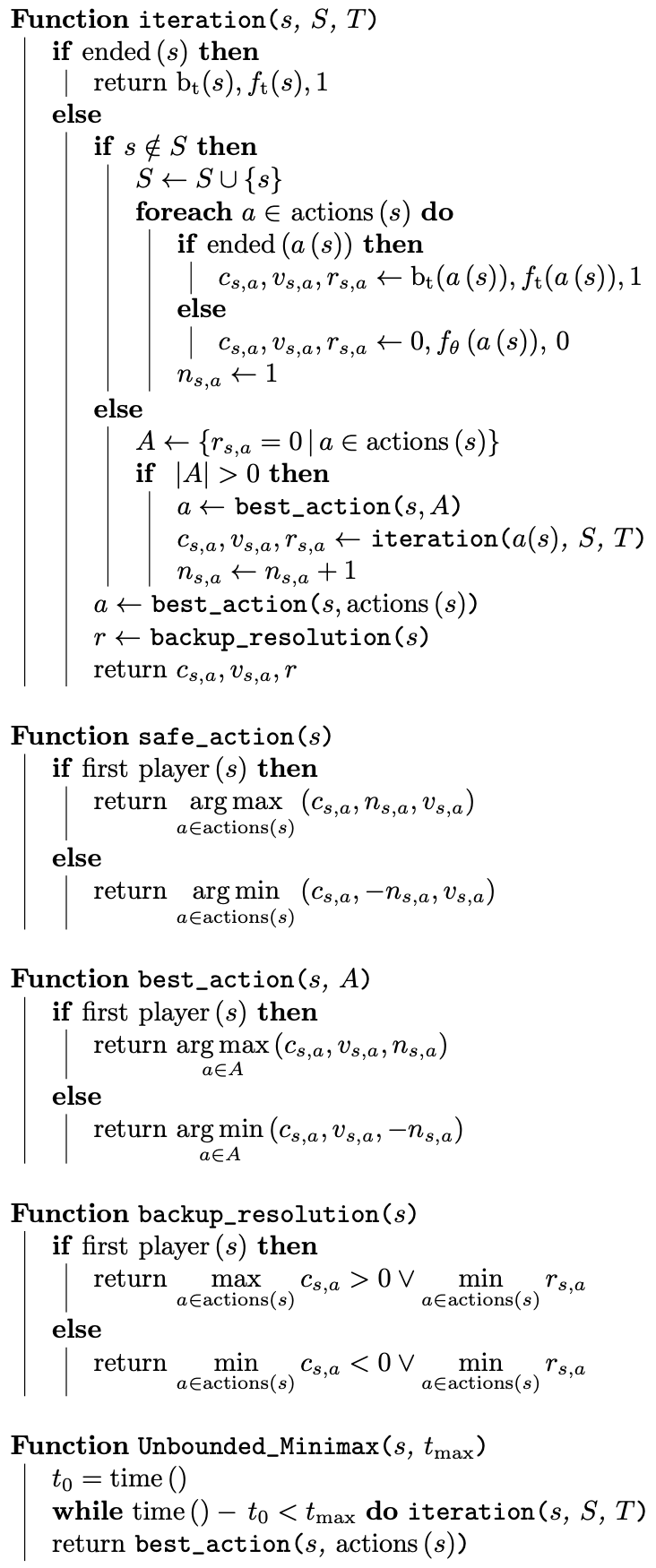}
\par\end{centering}
\caption{\textbf{Unbounded Best-First Minimax with completion.} It computes
the best action to play in the generated non-uniform partial game
tree starting from the root state $s$ with $\protect\tmax$ as research
time. See Table~\ref{tab:Index-of-symbols} for the definitions of
symbols. Note: tuples are lexicographically ordered. To obtain the
safe decision version: replace best\_action in the last line of Unbounded\_Minimax
by safe\_action.}\label{alg:ubfms}
\end{algorithm}

\section{Contributions}

\subsection{A New Search Algorithm for Multi-Player Games}\label{subsec:Unbounded_maxn}

In this section, we introduce our generalization of Unbounded Minimax
to the multi-player setting. We call this algorithm Unbounded $\maxn$.
Like Unbounded Minimax, this algorithm iteratively extends the a priori
best action sequence as long as search time remains. The difference
is that, state values are now tuples of values \LyXZeroWidthSpace corresponding
to each player's values and instead of the first player maximizing
the terminal value and the second player minimizing it, each player
here independently maximizes their own value at its turn (player $p$
maximizes its completion value $c(s,a)_{p}$ and its partial minimax
value $v(s,a)_{p}$). Thus, in this generalized context, the notion
of a “best” action reduces to maximizing the current player’s own
utility component, independently of the other players’ outcomes (this
is not necessarily optimal, since the actual objective is to achieve
a higher payoff than one’s opponents rather than to maximize an absolute
score).

The Unbounded $\maxn$ algorithm with completion and optionally with
safe decision is described in Algorithm~\ref{alg:umaxn}. 

\begin{algorithm}
\begin{centering}
\includegraphics[scale=0.5]{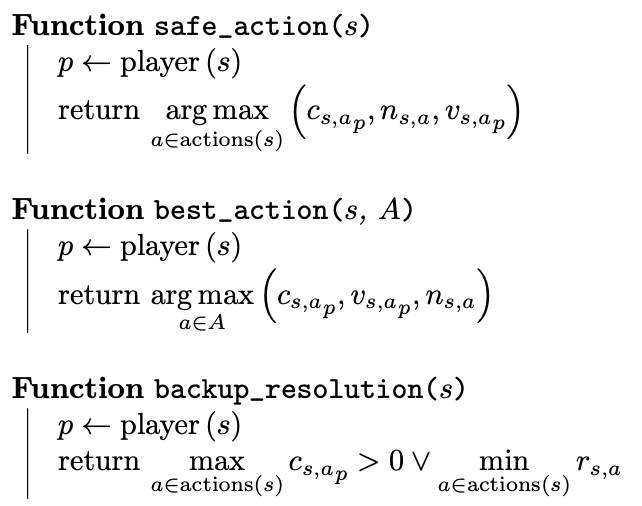}
\par\end{centering}
\caption{\textbf{Unbounded $\protect\maxn$ (with completion).} The Unbounded\_Minimax
method (which should therefore be renamed here as Unbounded\_$\protect\maxn$)
and iteration method are the same as in the two-player case (see Algorithm~\ref{alg:ubfms})
(except that $c(s,a)$ is initialized to $(0,\ldots,0)$ instead of
$0$). The other methods are overloaded and described here. This algorithm
computes the best action to play in the generated non-uniform partial
game tree starting from the root state $s$ with $\protect\tmax$
as research time. See Table~\ref{tab:Index-of-symbols} for the definitions
of symbols. Note: tuples are lexicographically ordered. To obtain
the safe decision version: replace best\_action in the last line of
Unbounded\_Minimax by safe\_action.}\label{alg:umaxn}
\end{algorithm}

\subsection{Experimental Study of Multi-Player Searches}\label{sec:Experiments}

We now present the experiments conducted to evaluate the contributions
of this paper. We will present its results in the following section. 

We begin by presenting the experimental protocol (Section~\ref{subsec:Evaluation-Protocol}).
Then we explain how the evaluation functions used by the search algorithms
were generated (Section~\ref{subsec:apprentissage-fonctions-evaluation}).
Next, we describe the machines used (Section~\ref{subsec:computer_used}),
and finally present the other details of the experiment (Section~\ref{subsec:Other-Details-of}).

\subsubsection{Evaluation Protocol}\label{subsec:Evaluation-Protocol}

We experimentally compare the following algorithms: $\brs$, $\brsp$,
$\mcts{}$, $\mctsh{}$, Paranoid, standard $\maxn$, $\maxn$ with
$k$-best pruning and our new algorithm Unbounded $\maxn$ with classic
decision and with safe decision.

We use standard $\maxn$ as benchmark adversary using a lossless parallelization
procedure, called Child Batching \cite{cohen2025study,cohen2020learning}
(child Batching consists in evaluating the children states of states
in parallel by the neural network). We evaluate each algorithm on
the following 8 games: 3 classic games: Hey, That's My Fish!, Blokus,
Three-Players Hex as well as 5 multiplayer versions of classic two-player
games that we introduce: Quadrothello and Triinversion (generalizing
the game of Othello to four and three player respectively), Quadamazons
(generalizing the game Amazons to four player), and other generalizations
of Hex with a different degree of interaction between players: Separed
Teamhex and Threehex. Corresponding rules are in Appendix.

Each algorithm faces the benchmark adversary $P$ times where $P$
is the number of players in the game: each evaluated algorithm takes
a different player number each time. In other words, it plays once
as the first player, once as the second player, etc... These matches
are repeated $30\times30$ times: we use 30 different evaluation functions
per game. More precisely, each algorithm is evaluated for each evaluation
fonction $f_{i}$ ($i\in\left\{ 0,29\right\} $) that it uses and
for each opponents evaluation function group $\left\{ f_{j+k}\right\} _{k=0}^{P-2}$
($j\in\left\{ 0,29\right\} $). In other words, for all $i,j\in\left\{ 0,29\right\} $,
the evaluated algorithm using $f_{i}$ confronts in a same match the
following opponents: the benchmark adversary with the evaluation $f_{j}$,
the benchmark adversary with the evaluation $f_{j+1}$, ..., and the
benchmark adversary with the evaluation $f_{j+P-2}$. This corresponds
to 2700 (resp. 3600) matches for each evaluated algorithm and for
each 3-player (resp. 4-player) game. The performance evaluation of
an algorithm for a specific game is therefore based on this number
of matches. The performance of a search algorithm is the average of
the binary scores obtained at the end of its matches (1 for a victory,
0 for a draw and -1 for a defeat). 

Each search algorithm uses the same search time: 10 seconds per action
(order of magnitude of search time in competition). 

In this experiment, we evaluated Unbounded $\maxn$ with safe decision
without Child Batching and the MTCS-based algorithms without Child
Batching because Child Batching cannot be used with $\mcts{}$. In
addition, we evaluated all other algorithms with Child Batching. 

Note that $\brs$ has not been evaluated on all games because that
would have required re-impeding all games to handle invalid states.
$\brs$ games are: Separed Teamhex, Quadrothello, 3 Player Hex, Threehex,
and Triinversion. 

Note that, regardless of whether the evaluated algorithms use Child
Batching, the benchmark adversary always does. Therefore, all algorithms
can be directly compared.

\subsubsection{Evaluation Functions Training}\label{subsec:apprentissage-fonctions-evaluation}

We now present how the evaluation functions used in the experiments
of this paper were learned.

The evaluation functions were generated using Athénan \cite{cohen2020learning}
; the reinforcement learning algorithm, by remplacing its core component
Descent Minimax, with its generalization for multi-player games~\cite{cohen2021completeness}
(which consists of replacing the minimax paradigm with the $\maxn$
paradigm). 

Athénan involves sequentially performing self-play matches using the
learned evaluation function to guide the search process, deciding
which actions to take during the matches and how to explore the search
tree. The data encountered (the associations states-values) is then
learned at the end of each match to update the evaluation function.
This thus forms a self-improving loop.

The Athénan hyperparameters used to generate each evaluation function
are the following: search time per action $t=2$, batch size $B=3000$,
duplication factor $\delta=3$, memory size $M=75$, the selection
action method is still the ordinal law but its parameter is $\sqrt{X}$
with $X$ a unform random variable in {[}0,1{]}, neural network optimizer:
Adam~\cite{kingma2014adam} with $\lambda=0.0001$ as learning rate.
The neural network architecture used is as follow: a convolutional
layer, $4$ residual blocks~\cite{he2016deep} (containing 2 convolutional
layers), a convolutional layer with $1\times1$ as kernel size, a
Global Sum Polling layer~\cite{aich2018global}, a flatten layer,
a dense layer with $N$ neurons (see Table in Appendix for the $N$
value), and a final dense layer with $P$ neurons (the output ; $P$:
the number of players). Each convolutional layer has $C$ channels
(see Table in Appendix for the $C$ value). The activation function
is the ReLU.

Each of the evaluation functions is the result of 10 days of training.

\subsubsection{Computer Used}\label{subsec:computer_used}

We provide details of the machines used in the experiments described
in this article.

For reinforcement learning the experiment evaluation functions, we
used the Jean-Zay A100 partition: 8 CPU with 58 Go of RAM and 1 GPU
A100 with 80 Go of RAM (per learning process).

To perform the evaluation matches, we used the Jean-Zay V100 partition:
10 CPU with 40 Go of RAM and 1 GPU V100 with 16 Go of RAM.

\subsubsection{Other Details of the Experiment}\label{subsec:Other-Details-of}

We give for each algorithm: the average performance over all games,
the associated 95\% stratified bootstrapping confidence interval (lower
and upper bound) and the performance for each game with its 95\% confidence
radius. The percentages detailing performance for each game have been
rounded to the nearest percent to reduce size and improve readability
of data tables.

Algorithms based on a fixed search depth ($\brs$, $\brsp$, $\maxn$
with and without $k$-best pruning, and Paranoid) use iterative deepening~\cite{korf1985depth}
to be parameterized by a search time. 

All algorithms use transposition tables~\cite{greenblatt1988greenblatt}.
This allows the game tree to be transformed into a directed acyclic
graph, which therefore simplifies the state space.

Evaluation functions for $\mctsh{}$ need to by

The evaluation functions for $\mctsh{}$ need to take values in $[0,1]$.
We use the following normalization, applied component by component,
to obtain such evaluation functions, transforming an evaluation function
$f$ with values in $\mathbb{R}$ into an evaluation function $f_{\mathrm{n}}$
with values in $[0,1]$:

\[
f_{\mathrm{n}}(s)=\frac{\max\left(\min\left(f\left(s\right),M\right),m\right)-m}{M-m}
\]
with $s$ a game state, $f$ the state evaluation function, $M$ the
maximum practical value for $f$, $m$ the minimum practical value
of $f$. More formally, $M=\max V$ and $m=\min V$ with $V=\liste{f\left(s'\right)}{s'\in S}$
where $S$ is a set of states from matches played using $f$ generated
during 24 hours.

\subsection{Results of the Experimental Study}\label{subsec:Results-of-the}

\begin{table*}[t]
\begin{centering}
{\tiny{}%
\begin{tabular}{|c|c|c|c|c|c|c|c|c|}
\hline 
 & {\tiny Quadamazons} & {\tiny Blokus} & {\tiny Hey, That's My Fish!} & {\tiny Separed Teamhex} & {\tiny Quadrothello} & {\tiny 3 Player Hex} & {\tiny Triinversion} & {\tiny Threehex}\tabularnewline
\hline 
\hline 
{\tiny Safe Unbounded $\maxn$} & {\tiny\textbf{\textcolor{orange}{{} -45 \textpm 3 }}} & {\tiny\textbf{\textcolor{red}{{} -46 \textpm 3 }}} & {\tiny\textbf{\textcolor{red}{{} 1 \textpm 4   }}} & {\tiny\textbf{\textcolor{red}{{} -16 \textpm 3 }}} & {\tiny\textbf{\textcolor{red}{{} -16 \textpm 3 }}} & {\tiny\textbf{\textcolor{orange}{{} -7 \textpm 1  }}} & {\tiny\textbf{\textcolor{red}{{} -1 \textpm 4  }}} & {\tiny\textbf{\textcolor{red}{{} -33 \textpm 4}}}\tabularnewline
\hline 
{\tiny$\maxn$} & {\tiny -46\% \textpm 2\% } & {\tiny{} -48\% \textpm 2\% } & {\tiny{} -27\% \textpm 3\% } & {\tiny{} -16\% \textpm 3\% } & {\tiny{} -53\% \textpm 2\% } & {\tiny{} -7\% \textpm 1\%  } & {\tiny{} -37\% \textpm 3\% } & {\tiny{} -36\% \textpm 3\%}\tabularnewline
\hline 
\hline 
{\tiny$\mctsh{C=\sqrt{2}}$} & {\tiny -54 \textpm 2} & {\tiny -79 \textpm 1} & {\tiny -58 \textpm 3} & {\tiny -38 \textpm 3} & {\tiny -79 \textpm 1} & {\tiny -7 \textpm 1} & {\tiny -70 \textpm 2} & {\tiny -43 \textpm 3}\tabularnewline
\hline 
{\tiny$\mctsh{C=\frac{\sqrt{2}}{2}}$} & {\tiny -54 \textpm 2} & {\tiny -78 \textpm 2} & {\tiny -55 \textpm 3} & {\tiny -37 \textpm 3} & {\tiny -75 \textpm 2} & {\tiny -8 \textpm 1} & {\tiny -67 \textpm 2} & {\tiny -45 \textpm 3}\tabularnewline
\hline 
{\tiny$\mctsh{C=\frac{\sqrt{2}}{4}}$} & {\tiny -54 \textpm 2} & {\tiny -76 \textpm 2} & {\tiny -48 \textpm 3} & {\tiny -39 \textpm 3} & {\tiny -68 \textpm 2} & {\tiny -8 \textpm 1} & {\tiny -64 \textpm 2} & {\tiny -44 \textpm 3}\tabularnewline
\hline 
{\tiny$\mctsh{C=\frac{\sqrt{2}}{8}}$} & {\tiny -56 \textpm 2} & {\tiny -76 \textpm 2} & {\tiny -42 \textpm 3} & {\tiny -36 \textpm 3} & {\tiny -64 \textpm 2} & {\tiny -7 \textpm 1} & {\tiny -57 \textpm 3} & {\tiny -45 \textpm 3}\tabularnewline
\hline 
\hline 
{\tiny$\mcts{C=\sqrt{2}}$} & {\tiny -28 \textpm 3} & {\tiny -99 \textpm 0} & {\tiny -92 \textpm 1} & {\tiny -92 \textpm 1} & {\tiny -100 \textpm 0} & {\tiny -1 \textpm 0} & {\tiny -85 \textpm 1} & {\tiny -97 \textpm 0}\tabularnewline
\hline 
{\tiny$\mcts{C=\frac{\sqrt{2}}{2}}$} & {\tiny -20 \textpm 3} & {\tiny -99 \textpm 0} & {\tiny -90 \textpm 1} & {\tiny -91 \textpm 1} & {\tiny -100 \textpm 0} & {\tiny 0 \textpm 0} & {\tiny -86 \textpm 1} & {\tiny -96 \textpm 1}\tabularnewline
\hline 
{\tiny$\mcts{C=\frac{\sqrt{2}}{4}}$} & {\tiny\textbf{\textcolor{red}{-11 \textpm 3}}} & {\tiny -99 \textpm 0} & {\tiny -87 \textpm 1} & {\tiny -88 \textpm 1} & {\tiny -100 \textpm 0} & {\tiny 1 \textpm 0} & {\tiny -84 \textpm 2} & {\tiny -94 \textpm 1}\tabularnewline
\hline 
{\tiny$\mcts{C=\frac{\sqrt{2}}{8}}$} & {\tiny -16 \textpm 3} & {\tiny -99 \textpm 0} & {\tiny -88 \textpm 1} & {\tiny -88 \textpm 1} & {\tiny -100 \textpm 0} & {\tiny\textbf{\textcolor{red}{2 \textpm 0}}} & {\tiny -86 \textpm 1} & {\tiny -92 \textpm 1}\tabularnewline
\hline 
\end{tabular}}{\tiny}{\tiny\par}
\par\end{centering}
\caption{Binary scores for all games of the evaluated algorithms without batching
against $\protect\maxn$ with batching (\textcolor{red}{red} \textgreater{}
\textcolor{orange}{orange} \textgreater{} \textcolor{pink}{pink}).}\label{tab:res_sans-batch-details}
\end{table*}

We now present the results of the experiment. In Section~\ref{subsec:First-stage-of},
we present the performance of $\mcts{}$ and $\mctsh{}$ as well as
some algorithms without Child Batching. Section~\ref{subsec:Second-stage-of}
presents the performance of $\brs$ by comparing it with other algorithms
using also Child Batching, restricting our analysis to the games where
$\brs$ can be applied. In Section~\ref{subsec:Algorithms-with-Child},
we present the performance of all the algorithms, except $\mcts{}$,
$\mctsh{}$, and $\brs$, with Child Batching, on all the games. 

\subsubsection{$\protect\mcts{}$, $\protect\mctsh{}$, and algorithms without Child
Batching}\label{subsec:First-stage-of}

We thus evaluate the following algorithms without Child Batching:
$\mcts{}$, $\mctsh{}$, and Unbounded $\maxn$ with Safe Decision
without Child Batching, and $\maxn$. The algorithms performances
are described in Table~\ref{tab:res_sans-batch-mean} (mean over
the games) and Table~\ref{tab:res_sans-batch-details} (details).

Safe Unbounded $\maxn$ achieves the best average performance, followed
by $\maxn$, then $\mctsh{}$, and finally $\mcts{}$. This result
generalizes across all games except Quadamazons and 3-Player Hex.
On these two games, $\mcts{}$ achieves the best performance, after
which the average performance ranking is recovered. Notably, $\mcts{}$
even outperforms the algorithms using Child Batching on Quadamazons
and equalizes them on 3-Player Hex (see Table~\ref{tab:res_avec-batch-details}).
Since $\mcts{}$ does not rely on any game-specific knowledge, we
infer that the learned evaluation functions for these two games are
of low quality.

The first conclusion of this paper, which is not surprising, is that
when learning fails to provide a high-quality evaluation function,
$\mcts{}$ delivers better performance. This study nevertheless shows
that such cases are rather rare (one game out of eight which is problematic).

Note that Child Batching increases performance by about 6\%.

Let us now focus more specifically on $\mctsh{}$. Excluding $\mcts{}$,
$\mctsh{}$ achieves the worst performance on all games. When compared
to the other algorithms using Child Batching (see Table~\ref{tab:res_avec-batch-details}),
$\mctsh{}$ remains far inferior, even after removing the 6\% gain
due to Child Batching, and even when considering the best parameters
for $\mcts{}$ and the worst parameters for the other algorithms.
We note a single exception: $\mctsh{}$ performs better than Paranoid
on the sole game Quadrothello, which can be interpreted as the fact
that assuming all opponents seek to minimize one's score is a very
poor heuristic for this game, despite the computational gains.

The second conclusion is thus that $\mctsh{}$ is the worst-performing
algorithm in the context of our experiments, even if it could be used
with Child Batching.

\begin{table}
\begin{centering}
{\small{}%
\begin{tabular}{|c|c|c|c|}
\hline 
 & {\small mean} & {\small lower bound} & {\small upper bound}\tabularnewline
\hline 
\hline 
{\small Safe Unbounded $\maxn$} & {\small\textbf{\textcolor{red}{-21.88 }}} & {\small -22.83 } & {\small -20.94}\tabularnewline
\hline 
{\small$\maxn$} & {\small -35.23 } & {\small -36.13 } & {\small -34.34}\tabularnewline
\hline 
{\small$\mctsh{}$ (best $C=\frac{\sqrt{2}}{8}$)} & {\small\textbf{\textcolor{orange}{-49.76}}} & {\small -50.6} & {\small -48.96}\tabularnewline
\hline 
{\small$\mcts{}$ (best $C=\frac{\sqrt{2}}{4}$)} & {\small\textbf{\textcolor{pink}{-71.27}}} & {\small -71.79} & {\small -70.76}\tabularnewline
\hline 
\end{tabular}}{\small}{\small\par}
\par\end{centering}
\caption{Average binary scores over all games of the evaluated algorithms without
batching against $\protect\maxn$ with batching.}\label{tab:res_sans-batch-mean}
\end{table}

\subsubsection{$\protect\brs$ with Child Batching}\label{subsec:Second-stage-of}

We now analyze the performance of $\brs$ by comparing it with the
other algorithms while restricting ourselves to the games on which
$\brs$ could be applied. Recall that $\brs$ creates invalid states
that are not allowed by the game rules, and therefore it requires
modifying the game implementations in order to be used, which entails
a far from negligible human cost. 

Performance of $\brs$ are described in Table~\ref{tab:res_avec-batch-mean-brs}
(mean over the $\brs$ games) and Table~\ref{tab:res_avec-batch-details}
(details). $\brs$ is the algorithm with the worst average performance
among the algorithms using child batching. In particular, it achieves
the poorest performance on each of the two generalizations of Othello.
However, it belongs to the leading group on the three generalizations
of Hex. Nevertheless, $\brsp$ consistently obtains better results
than $\brs$. This study therefore confirms the inferiority of $\brs$
compared to $\brsp$ and indicates that there is no need to further
investigate $\brs$.

\begin{table}
\begin{centering}
{\small{}%
\begin{tabular}{|c||c|c|c|}
\hline 
 & {\small mean} & {\small lower bound} & {\small upper bound}\tabularnewline
\hline 
\hline 
{\small Safe Unbounded $\maxn$} & {\small\textbf{\textcolor{red}{-8.25}}} & {\small -9.46} & {\small -7.03}\tabularnewline
\hline 
{\small Unbounded $\maxn$} & {\small\textbf{\textcolor{orange}{-24.41}}} & {\small -25.57} & {\small -23.27}\tabularnewline
\hline 
{\small$\brs$} & {\small -39.15} & {\small -40.14} & {\small -38.11}\tabularnewline
\hline 
{\small$\brsp$} & {\small -32.03} & {\small -33.12} & {\small -30.95}\tabularnewline
\hline 
{\small Paranoid} & {\small -32.6} & {\small -33.67} & {\small -31.54}\tabularnewline
\hline 
{\small$\maxn$} & {\small\textbf{\textcolor{pink}{-26.86}}} & {\small -28.01} & {\small -25.71}\tabularnewline
\hline 
\end{tabular}}{\small}{\small\par}
\par\end{centering}
\caption{Average binary scores over the games where $\protect\brs$ has been
used of the evaluated algorithms with batching against $\protect\maxn$
with batching (\textcolor{red}{red} \textgreater{} \textcolor{orange}{orange}
\textgreater{} \textcolor{pink}{pink}).}\label{tab:res_avec-batch-mean-brs}
\end{table}

\begin{table*}
\begin{centering}
{\tiny{}%
\begin{tabular}{|c||c|c|c|c|c|c|c|c|}
\hline 
 & {\tiny Quadamazons} & {\tiny Blokus} & {\tiny Hey, That's My Fish!} & {\tiny Separed Teamhex} & {\tiny Quadrothello} & {\tiny 3 Player Hex} & {\tiny Triinversion} & {\tiny Threehex}\tabularnewline
\hline 
\hline 
{\tiny Safe Unbounded $\maxn$} & {\tiny\textbf{\textcolor{red}{-40 \textpm 3 }}} & {\tiny\textbf{\textcolor{pink}{{} -45 \textpm 3 }}} & {\tiny\textbf{\textcolor{red}{{} 6 \textpm 4   }}} & {\tiny\textbf{\textcolor{red}{{} -10 \textpm 3 }}} & {\tiny\textbf{\textcolor{red}{{} -12 \textpm 3 }}} & {\tiny\textbf{\textcolor{red}{{} 1 \textpm 1   }}} & {\tiny\textbf{\textcolor{red}{{} 5 \textpm 4   }}} & {\tiny\textbf{\textcolor{red}{{} -24 \textpm 4}}}\tabularnewline
\hline 
{\tiny Unbounded $\maxn$} & {\tiny -45 \textpm 2} & {\tiny -58 \textpm 2} & {\tiny\textbf{\textcolor{pink}{-13 \textpm 3}}} & {\tiny -18 \textpm 3} & {\tiny\textbf{\textcolor{orange}{-33 \textpm 3}}} & {\tiny\textbf{\textcolor{pink}{-2 \textpm 1}}} & {\tiny\textbf{\textcolor{orange}{-19 \textpm 3}}} & {\tiny -46 \textpm 3}\tabularnewline
\hline 
{\tiny$\brs$} &  &  &  & {\tiny -15 \textpm 3} & {\tiny -89 \textpm 1} & {\tiny -1 \textpm 1} & {\tiny -54 \textpm 3} & {\tiny -25 \textpm 3}\tabularnewline
\hline 
{\tiny$\brsp$} & {\tiny\textbf{\textcolor{red}{-37 \textpm 3}}} & {\tiny\textbf{\textcolor{orange}{-42 \textpm 2}}} & {\tiny\textbf{\textcolor{pink}{-15 \textpm 3}}} & {\tiny\textbf{\textcolor{orange}{-12 \textpm 3}}} & {\tiny -81 \textpm 1} & {\tiny\textbf{\textcolor{orange}{-1 \textpm 1}}} & {\tiny\textbf{\textcolor{pink}{-29 \textpm 3}}} & {\tiny\textbf{\textcolor{red}{-24 \textpm 3}}}\tabularnewline
\hline 
{\tiny Paranoid} & {\tiny\textbf{\textcolor{red}{-39 \textpm 3}}} & {\tiny\textbf{\textcolor{red}{-37 \textpm 3}}} & {\tiny -25 \textpm 3} & {\tiny\textbf{\textcolor{red}{-9 \textpm 3}}} & {\tiny -81 \textpm 1} & {\tiny\textbf{\textcolor{orange}{-1 \textpm 1}}} & {\tiny -36 \textpm 3} & {\tiny\textbf{\textcolor{red}{-24 \textpm 3}}}\tabularnewline
\hline 
{\tiny$\maxn$} & {\tiny -45 \textpm 2} & {\tiny\textbf{\textcolor{pink}{-47 \textpm 2}}} & {\tiny -28 \textpm 3} & {\tiny\textbf{\textcolor{orange}{-13 \textpm 3}}} & {\tiny\textbf{\textcolor{pink}{-48 \textpm 2}}} & {\tiny -4 \textpm 1} & {\tiny\textbf{\textcolor{pink}{-33 \textpm 3}}} & {\tiny -32 \textpm 3}\tabularnewline
\hline 
\hline 
{\tiny$k$-best $\maxn$ ($k=30$)} & {\tiny -44 \textpm 2} & {\tiny -47 \textpm 2} & {\tiny -29 \textpm 3} & {\tiny -14 \textpm 3} & {\tiny -50 \textpm 2} & {\tiny -4 \textpm 1} & {\tiny -35 \textpm 3} & {\tiny -32 \textpm 3}\tabularnewline
\hline 
{\tiny$k$-best $\maxn$ ($k=20$)} & {\tiny -44 \textpm 2} & {\tiny -47 \textpm 2} & {\tiny -25 \textpm 3} & {\tiny -12 \textpm 3} & {\tiny -50 \textpm 2} & {\tiny -4 \textpm 1} & {\tiny -39 \textpm 3} & {\tiny -35 \textpm 3}\tabularnewline
\hline 
{\tiny$k$-best $\maxn$ ($k=14$)} & {\tiny -43 \textpm 2} & {\tiny -48 \textpm 2} & {\tiny -21 \textpm 3} & {\tiny -14 \textpm 3} & {\tiny -50 \textpm 2} & {\tiny -3 \textpm 1} & {\tiny -41 \textpm 3} & {\tiny -34 \textpm 3}\tabularnewline
\hline 
{\tiny$k$-best $\maxn$ ($k=10$)} & {\tiny -45 \textpm 2} & {\tiny -46 \textpm 2} & {\tiny -12 \textpm 3} & {\tiny -13 \textpm 3} & {\tiny -45 \textpm 2} & {\tiny -3 \textpm 1} & {\tiny -34 \textpm 3} & {\tiny -34 \textpm 3}\tabularnewline
\hline 
{\tiny$k$-best $\maxn$ ($k=8$)} & {\tiny -47 \textpm 2} & {\tiny -50 \textpm 2} & {\tiny -16 \textpm 3} & {\tiny -16 \textpm 3} & {\tiny -46 \textpm 2} & {\tiny -3 \textpm 1} & {\tiny -36 \textpm 3} & {\tiny -34 \textpm 3}\tabularnewline
\hline 
{\tiny$k$-best $\maxn$ ($k=7$)} & {\tiny -46 \textpm 2} & {\tiny -52 \textpm 2} & {\tiny -13 \textpm 3} & {\tiny -13 \textpm 3} & {\tiny -42 \textpm 2} & {\tiny -3 \textpm 1} & {\tiny -28 \textpm 3} & {\tiny -32 \textpm 3}\tabularnewline
\hline 
{\tiny$k$-best $\maxn$ ($k=6$)} & {\tiny -46 \textpm 2} & {\tiny -52 \textpm 2} & {\tiny -10 \textpm 3} & {\tiny -14 \textpm 3} & {\tiny -40 \textpm 2} & {\tiny -3 \textpm 1} & {\tiny -33 \textpm 3} & {\tiny -30 \textpm 3}\tabularnewline
\hline 
{\tiny$k$-best $\maxn$ ($k=5$)} & {\tiny -48 \textpm 2} & {\tiny -56 \textpm 2} & {\tiny -14 \textpm 3} & {\tiny -15 \textpm 3} & {\tiny -34 \textpm 3} & {\tiny -2 \textpm 1} & {\tiny -29 \textpm 3} & {\tiny -32 \textpm 3}\tabularnewline
\hline 
{\tiny$k$-best $\maxn$ ($k=4$)} & {\tiny -46 \textpm 2} & {\tiny -61 \textpm 2} & {\tiny\textbf{\textcolor{orange}{-8 \textpm 3}}} & {\tiny -17 \textpm 3} & {\tiny\textbf{\textcolor{orange}{-33 \textpm 3}}} & {\tiny\textbf{\textcolor{pink}{-2 \textpm 1}}} & {\tiny -24 \textpm 3} & {\tiny -32 \textpm 3}\tabularnewline
\hline 
{\tiny$k$-best $\maxn$ ($k=3$)} & {\tiny -44 \textpm 2} & {\tiny -62 \textpm 2} & {\tiny -9 \textpm 3} & {\tiny -21 \textpm 3} & {\tiny -34 \textpm 3} & {\tiny -4 \textpm 1} & {\tiny\textbf{\textcolor{orange}{-20 \textpm 3}}} & {\tiny -31 \textpm 3}\tabularnewline
\hline 
{\tiny$k$-best $\maxn$ ($k=2$)} & {\tiny -50 \textpm 2} & {\tiny -69 \textpm 2} & {\tiny -12 \textpm 3} & {\tiny -20 \textpm 3} & {\tiny -39 \textpm 3} & {\tiny -6 \textpm 1} & {\tiny -35 \textpm 3} & {\tiny -33 \textpm 3}\tabularnewline
\hline 
\end{tabular}}{\tiny}{\tiny\par}
\par\end{centering}
\caption{Binary scores for all games of the evaluated algorithms with batching
against $\protect\maxn$ with batching (\textcolor{red}{red} \textgreater{}
\textcolor{orange}{orange} \textgreater{} \textcolor{pink}{pink})}\label{tab:res_avec-batch-details}
\end{table*}

\subsubsection{Algorithms with Child Batching}\label{subsec:Algorithms-with-Child}

We now present the performance of the following algorithms with Child
Batching: $\brsp$, Paranoid, $\maxn$, $\maxn$ with $k$-best pruning,
and our new algorithms: Unbounded $\maxn$ with classic decision and
Unbounded $\maxn$ with safe decision. 

The algorithms performances are described in Table~\ref{tab:res_avec-batch-mean}
(mean over the games) and Table~\ref{tab:res_avec-batch-details}
(details). 

Unbounded $\maxn$ with safe decision obtains the best average performance
and it is the best algorithm, possibly tied, over $7$ of the $8$
studied games (Blokus is the exception). It is strictly the best algorithm
over $3$ of the $8$ studied games (Quadrothello, Triinversion, Threehex).

Unbounded $\maxn$ without Safe Decision obtains the second best average
performance. However, it is not the best-performing algorithm on any
game and achieves second-best performance on only 3 games. On 4 of
the 8 games, it obtains performance that is significantly worse than
$\brsp$, Paranoid, and $\maxn$.

In addition, $\maxn$ and $\brsp$ achieve the third-best average
performance. $\brsp$ achieves the best performance on 2 of the 8
games (Quadamazons and Threehex). $\maxn$ is not the best-performing
algorithm on any game. In fact, it is never better than $\brsp$,
except on a single game, Quadrothello, where it performs largely better.

Moreover, Paranoid is the fifth-best algorithm in terms of average
performance. Nevertheless, it reaches the best performance level on
4 of the 8 games (Quadamazons, Blokus, Separed Teamhex, Threehex).

Finally, we note that $k$-best pruning improves performance of $\maxn$
on only 3 of the 8 games (Hey, That's My Fish!, Quadrothello, Triinversion).
$\brsp$ when combined with $k$-best pruning could possibly match
the performance of Safe Unbounded $\maxn$ on one of the 3 games (namely,
Hey, That's My Fish!). For the other two games, using $k$-best pruning
with $\brsp$ or Paranoid should not change the relative performance
ranking of the algorithms.

\begin{table}
\begin{centering}
{\small{}%
\begin{tabular}{|c||c|c|c|}
\hline 
 & {\small mean} & {\small lower bound} & {\small upper bound}\tabularnewline
\hline 
\hline 
{\small Safe Unbounded $\maxn$} & {\small\textbf{\textcolor{red}{-16.48}}} & {\small -17.42} & {\small -15.54}\tabularnewline
\hline 
{\small Unbounded $\maxn$} & {\small\textbf{\textcolor{orange}{-31.05}}} & {\small -31.96} & {\small -30.13}\tabularnewline
\hline 
{\small$\brsp$} & {\small\textbf{\textcolor{pink}{-32.61}}} & {\small -33.5} & {\small -31.71}\tabularnewline
\hline 
{\small Paranoid} & {\small -33.46} & {\small -34.35} & {\small -32.58}\tabularnewline
\hline 
{\small$\maxn$} & {\small\textbf{\textcolor{pink}{-32.69}}} & {\small -33.6} & {\small -31.77}\tabularnewline
\hline 
{\small$k$-best $\maxn$ (best $k=6$)} & {\small\textbf{\textcolor{orange}{-30.39}}} & {\small -31.31} & {\small -29.48}\tabularnewline
\hline 
\end{tabular}}{\small}{\small\par}
\par\end{centering}
\caption{Average binary scores over all games of the evaluated algorithms with
batching against $\protect\maxn$ with batching.}\label{tab:res_avec-batch-mean}
\end{table}

\section{Conclusion}\label{sec:Conclusion}

\subsection{Summary}

In this paper, we investigated search algorithms for multi-player
perfect-information games. We generalized the state-of-the-art algorithm
for two-player zero-sum perfect-information games, namely Unbounded
Minimax with Safe Decision, and its variant: Unbounded Minimax with
classic decision, to this setting.

We then compared their performances in an exceptionally large and
robust study with the main multi-player search algorithms from the
literature. We showed that our generalization with safe decision also
achieves the best average performance for a medium search time. More
precisely, we have shown that it is the best-performing algorithm
for 7 out of the 8 games studied, and significantly superior in 3
games.

In addition, we retrieve several findings from the literature, but
this time within a robust and large study: Paranoid and $\brsp$ which
excel on certain games, $k$-best pruning which improves performance
on some games, $\brsp$ which outperforms $\brs$, and safe decision
which outperforms classic decision.

More precisely, we showed that our generalization of Unbounded Minimax
with classic decision achieves the second-best average performance
but is never the top-performing algorithm on any individual game.

We also repeated this study in the context of a short search time
(1 second), the results, available in Appendix, are analogous to the
medium times (10 seconds).

Overall, we have shown that our algorithm, Safe Unbounded $\maxn$,
is the search algorithm to be preferred as a first choice for multi-player
perfect-information games, when learning an evaluation function is
feasible (i.e., when access to a GPU over several days is available).
When this is not possible, or in the rare cases where learning fails
to provide a sufficiently strong evaluation function, base MCTS is
the algorithm of choice.

\subsection{Future work}

On the game where Unbounded $\maxn$ is not the best, namely Blokus,
Paranoid achieves the best result. A hybridization of Paranoid with
Unbounded $\maxn$, that is, modeling Blocus as a two-player game
using the Paranoid strategy allowing the application of Safe Unbounded
Minimax (the two-player algorithm), should constitute the state of
the art for this game. This hypothesis will be verified in a future
study. More generally, this modification could represent the state
of the art on four other games, since Safe Unbounded $\maxn$ matches
Paranoid performance on these games. 

More generally, we plan to study how the degree of interaction between
players affects the performance of search algorithms, and if performance
is linked, design a search algorithm that automatically and dynamically
adapts during the game to the current degree of interaction to optimize
its performance.

\clearpage{}

\bibliographystyle{plain}
\bibliography{jeux}

@article{baier2018mcts,
  title={MCTS-Minimax Hybrids with State Evaluations},
  author={Baier, Hendrik and Winands, Mark HM},
  journal={Journal of Artificial Intelligence Research},
  volume={62},
  pages={193--231},
  year={2018}
}

@inproceedings{ramanujan2011trade,
  title={Trade-Offs in Sampling-Based Adversarial Planning.},
  author={Ramanujan, Raghuram and Selman, Bart},
  booktitle={ICAPS},
  pages={202--209},
  year={2011}
}

@article{browne2012survey,
  title={A survey of monte carlo tree search methods},
  author={Browne, Cameron B and Powley, Edward and Whitehouse, Daniel and Lucas, Simon M and Cowling, Peter I and Rohlfshagen, Philipp and Tavener, Stephen and Perez, Diego and Samothrakis, Spyridon and Colton, Simon},
  journal={Transactions on Computational Intelligence and AI in games},
  volume={4},
  number={1},
  pages={1--43},
  year={2012},
  publisher={IEEE}
}

@incollection{greenblatt1988greenblatt,
  title={The Greenblatt chess program},
  author={Greenblatt, Richard D and Eastlake, Donald E and Crocker, Stephen D},
  booktitle={Computer chess compendium},
  pages={56--66},
  year={1988},
  publisher={Springer}
}

@article{korf1985depth,
  title={Depth-first iterative-deepening: An optimal admissible tree search},
  author={Korf, Richard E},
  journal={Artificial Intelligence},
  volume={27},
  number={1},
  pages={97--109},
  year={1985},
  publisher={Elsevier}
}

@article{korf1996best,
  title={Best-first minimax search},
  author={Korf, Richard E and Chickering, David Maxwell},
  journal={Artificial intelligence},
  volume={84},
  number={1-2},
  pages={299--337},
  year={1996},
  publisher={Elsevier}
}

@article{kingma2014adam,
  title={Adam: A method for stochastic optimization},
  author={Kingma, Diederik P and Ba, Jimmy},
  journal={arXiv preprint arXiv:1412.6980},
  year={2014}
}

@inproceedings{he2016deep,
  title={Deep residual learning for image recognition},
  author={He, Kaiming and Zhang, Xiangyu and Ren, Shaoqing and Sun, Jian},
  booktitle={Conference on Computer Vision and Pattern Recognition},
  pages={770--778},
  year={2016}
}

@article{knuth1975analysis,
  title={An analysis of alpha-beta pruning},
  author={Knuth, Donald E and Moore, Ronald W},
  journal={Artificial Intelligence},
  volume={6},
  number={4},
  pages={293--326},
  year={1975},
  publisher={Elsevier}
}

@inproceedings{Coulom06,
  author    = {R{\'{e}}mi Coulom},
  title     = {Efficient Selectivity and Backup Operators in Monte-Carlo Tree Search},
  booktitle = {Computers and Games, 5th International Conference, {CG} 2006, Turin,
               Italy, May 29-31, 2006. Revised Papers},
  pages     = {72--83},
  year      = {2007}
}

@article{silver2018general,
  title={A general reinforcement learning algorithm that masters chess, shogi, and Go through self-play},
  author={Silver, David and Hubert, Thomas and Schrittwieser, Julian and Antonoglou, Ioannis and Lai, Matthew and Guez, Arthur and Lanctot, Marc and Sifre, Laurent and Kumaran, Dharshan and Graepel, Thore and others},
  journal={Science},
  volume={362},
  number={6419},
  pages={1140--1144},
  year={2018},
  publisher={American Association for the Advancement of Science}
}

@article{
cohen2020learning, title={Learning to Play Two-Player Perfect-Information Games without Knowledge}, author={Cohen-Solal, Quentin}, journal={arXiv preprint arXiv:2008.01188}, year={2020} }

@article{cohen2025study, 
title={Study and improvement of search algorithms in two-players perfect information games}, 
author={Cohen-Solal, Quentin}, 
journal={arXiv preprint arXiv:2505.09639}, 
year={2025} 
}

@article{cohen2025little, 
title={On some improvements to Unbounded Minimax}, 
author={Cohen-Solal, Quentin and Cazenave, Tristan}, 
journal={arXiv preprint arXiv:2505.04525}, 
year={2025} 
}

@article{sturtevant2000pruning,
  title={On pruning techniques for multi-player games},
  author={Sturtevant, Nathan R and Korf, Richard E},
  journal={AAAI/IAAI},
  volume={49},
  pages={201--207},
  year={2000}
}

@article{cohen2023minimax,
  title={Minimax Strikes Back},
  author={Cohen-Solal, Quentin and Cazenave, Tristan},
  journal={AAMAS},
  year={2023}
}

@article{cohen2021descent,
	title={DESCENT wins five gold medals at the Computer Olympiad},
	author={Cohen-Solal, Quentin and Cazenave, Tristan},
	journal={ICGA Journal},
	volume={43},
	number={2},
	pages={132--134},
	year={2021},
	publisher={IOS Press}
}

@article{cohen2023athenan,
	title={Athenan Wins Sixteen Gold Medals at the Computer Olympiad},
	author={Cohen-Solal, Quentin and Cazenave, Tristan},
	journal={ICGA Journal},
	volume={45},
	number={3},
	year={2023},
}

@article{chao2018blokus,
  title={Blokus Game Solver},
  author={Chao, Chin},
  year={2018}
}

@article{aich2018global,
  title={Global sum pooling: A generalization trick for object counting with small datasets of large images},
  author={Aich, Shubhra and Stavness, Ian},
  journal={arXiv preprint arXiv:1805.11123},
  year={2018}
}

@inproceedings{luckhart1986algorithmic,
  title={An Algorithmic Solution of N-Person Games.},
  author={Luckhart, Carol and Irani, Keki B},
  booktitle={AAAI},
  volume={86},
  pages={158--162},
  year={1986}
}

@article{schadd2011best,
  title={Best reply search for multiplayer games},
  author={Schadd, Maarten PD and Winands, Mark HM},
  journal={IEEE Transactions on Computational Intelligence and AI in Games},
  volume={3},
  number={1},
  pages={57--66},
  year={2011},
  publisher={IEEE}
}

@article{cohen2021completeness,
  title={Completeness of Unbounded Best-First Game Algorithms},
  author={Cohen-Solal, Quentin},
  journal={arXiv preprint arXiv:2109.09468},
  year={2021}
}

@article{cohen2025athenan,
	title={Ath{\'e}nan Wins 11 Gold Medals at the 2024 Computer Olympiad},
	author={Cohen-Solal, Quentin and Cazenave, Tristan},
	journal={ICGA Journal},
	pages={13896911251315102},
	year={2025},
	publisher={SAGE Publications Sage UK: London, England}
}

@inproceedings{sturtevant2008analysis,
  title={An analysis of UCT in multi-player games},
  author={Sturtevant, Nathan R},
  booktitle={International conference on computers and games},
  pages={37--49},
  year={2008},
  organization={Springer}
}

@mastersthesis{hartjes2019feasibility,
  title={The Feasibility of Ignoring Opponents in Multi-Player Games},
  author={Hartjes, JO},
  year={2019}
}

@article{nijssen2013search,
  title={Search policies in multi-player games},
  author={Nijssen, JAM and Winands, Mark HM},
  journal={Icga Journal},
  volume={36},
  number={1},
  pages={3--21},
  year={2013},
  publisher={SAGE Publications Sage UK: London, England}
}

@article{nijssen2013monte,
  title={Monte-Carlo tree search for multi-player games},
  author={Nijssen, Joseph Antonius Maria},
  year={2013}
}

\clearpage{}

\section*{Appendix}

We present in this document the details of the experiments.

First, note that the $N$ value and $C$ value of the neural networks
for each game are in Table~\ref{tab:F_N_values_ExIt_Ath=0000E9nan}. 

In Section~\ref{subsec:Short-time-experiment}, we present the results
of our repeated experiment, modifying the associated search time to
analyze the short-time behavior of the algorithms.

In Section~\ref{subsec:Games}, we present the games used as benchmarks
for the experiments in this article.

\subsection{Short time experiment}\label{subsec:Short-time-experiment}

We repeated the experiment from our main article. Everything is identical
except for the search time of each evaluated algorithm and the reference
algorithm, which is set at 1 second (instead of 10 seconds).

Performance of algorithms without Child Batching is in Table~\ref{tab:res_sans-batch-mean-2}
(average) and in Table~\ref{tab:res_sans-batch-details-1} (details).
Average performance over all $\brs$ games of algorithms with Child
Batching is in Table~\ref{tab:res_avec-batch-mean-brs-1}. Performance
of algorithms with Child Batching is in Table~\ref{tab:res_avec-batch-mean-1}
(average) and in Table~\ref{tab:res_avec-batch-details-1} (details).

The results are analogous to the medium-time context, with a few minor
differences that we will detail.

$\maxn$ achieves better performance than Paranoid and $\brsp$. This
is interpreted by the fact that as the time is shorter, the impact
of the branching factor is less strong, there is therefore less need
for pruning, and therefore the pruning done by $\brsp$ and Paranoid
which are pruning with biases have a stronger negative impact on performance
(their bias is not compensated by the ability to plan relatively over
a longer term).

$k$-best $\maxn$ now significantly improves performance in Separated
Teamhex, and even achieves the best performance in this game, outclassing
Safe Unbounded $\maxn$. Although Unbounded $\maxn$ is still the
best algorithm on average, it is now only the best algorithm, possibly
tied, on 6 of the 8 games. However, in both games, it achieves performance
close to that of the best algorithm. In addition, it is now significantly
the best algorithm on 4 of the 8 games (the game Hey, That’s My Fish!
is added to the same list as for the 10-second time).

MCTS is no longer the best algorithm on Quadamazons. Therefore, MCTS
remains the worst algorithm in short time. We interpret this as meaning
that in short time, MCTS doesn't have enough time to converge to a
correct solution. Approaches based on learned heuristics are thus
more efficient because they begin the search already possessing some
knowledge.

$\mctsh{}$ is now better than Safe Unbounded $\maxn$ without Child
Batching on two games: Threehex and 3-Player Hex (and is still inferior
on the others). And we still have the fact that $\mctsh{}$ performs
worse than all other algorithms with child batching on all games except
two, instead of just one with 10 secondes: Quadrothello and Separated
Teamhex (Separated Teamhex is added). In these two exceptions, $\mctsh{}$
is still far from the best performance, especially on Quadrothello.

\begin{table}
\centering{}%
\begin{tabular}{|c|c|c|}
\hline 
 & $C$ & $N$\tabularnewline
\hline 
\hline 
Quadamazons & $166$ & 1343\tabularnewline
\hline 
Blokus & $166$ & 1260\tabularnewline
\hline 
Hey, That's My Fish! & $166$ & 1386\tabularnewline
\hline 
Separed Teamhex & $166$ & 1260\tabularnewline
\hline 
Quadrothello & $166$ & 1343\tabularnewline
\hline 
3-player Hex & $166$ & 1354\tabularnewline
\hline 
Triinversion & $166$ & 1373\tabularnewline
\hline 
Threehex & $166$ & 1354\tabularnewline
\hline 
\end{tabular}\caption{Number of convolutional channels $C$ and hidden dense neurons $N$
used by Athénan for each studied game.}\label{tab:F_N_values_ExIt_Ath=0000E9nan}
\end{table}

\begin{table}
\begin{centering}
{\small{}%
\begin{tabular}{|c|c|c|c|}
\hline 
 & {\small mean} & {\small lower bound} & {\small upper bound}\tabularnewline
\hline 
\hline 
{\small Safe Unbounded $\maxn$} & {\small\textbf{\textcolor{red}{-37.87}}} & {\small -38.77} & {\small -36.96}\tabularnewline
\hline 
{\small$\mctsh{C=\frac{\sqrt{2}}{4}}$} & {\small -47.63 } & {\small -48.46} & {\small -46.8}\tabularnewline
\hline 
{\small$\mctsh{C=\frac{\sqrt{2}}{8}}$} & {\small\textbf{\textcolor{orange}{-45.62 }}} & {\small -46.46} & {\small -44.77}\tabularnewline
\hline 
{\small$\mcts{C=\frac{\sqrt{2}}{4}}$} & {\small -85.12} & {\small -85.51} & {\small -84.74}\tabularnewline
\hline 
{\small$\mcts{C=\frac{\sqrt{2}}{8}}$} & {\small\textbf{\textcolor{pink}{-84.74}}} & {\small -85.12} & {\small -84.37}\tabularnewline
\hline 
\end{tabular}}{\small\par}
\par\end{centering}
\caption{Average binary scores over all games of the evaluated algorithms without
batching against $\protect\maxn$ with batching (search time $1s$
; \textcolor{red}{red} \textgreater{} \textcolor{orange}{orange} \textgreater{}
\textcolor{pink}{pink}).}\label{tab:res_sans-batch-mean-2}
\end{table}
\begin{table}
\begin{centering}
{\small{}%
\begin{tabular}{|c||c|c|c|}
\hline 
 & {\small mean} & {\small lower bound} & {\small upper bound}\tabularnewline
\hline 
\hline 
{\small Safe Unbounded $\maxn$} & {\small\textbf{\textcolor{red}{-21.1}}} & {\small -22.06} & {\small -20.13}\tabularnewline
\hline 
{\small Unbounded $\maxn$} & {\small\textbf{\textcolor{pink}{-30.69}}} & {\small -31.84} & {\small -29.51}\tabularnewline
\hline 
{\small$\brs$} & {\small -40.07} & {\small -41.1} & {\small -39.06}\tabularnewline
\hline 
{\small$\brsp$} & {\small -36.06} & {\small -37.14} & {\small -34.98}\tabularnewline
\hline 
{\small Paranoid} & {\small -36.65 } & {\small -37.71} & {\small -35.59}\tabularnewline
\hline 
{\small$\maxn$} & {\small\textbf{\textcolor{orange}{-28.24 }}} & {\small -29.39} & {\small -27.1}\tabularnewline
\hline 
\end{tabular}}{\small\par}
\par\end{centering}
\caption{Average binary scores over the games where $\protect\brs$ has been
used of the evaluated algorithms with batching against $\protect\maxn$
with batching (search time $1s$ ; \textcolor{red}{red} \textgreater{}
\textcolor{orange}{orange} \textgreater{} \textcolor{pink}{pink}).}\label{tab:res_avec-batch-mean-brs-1}
\end{table}

\begin{table*}[t]
\begin{centering}
{\tiny{}%
\begin{tabular}{|c|c|c|c|c|c|c|c|c|}
\hline 
 & {\tiny Quadamazons} & {\tiny Blokus} & {\tiny Hey, That's My Fish!} & {\tiny Separed Teamhex} & {\tiny Quadrothello} & {\tiny 3 Player Hex} & {\tiny Triinversion} & {\tiny Threehex}\tabularnewline
\hline 
\hline 
{\tiny Safe Unbounded $\maxn$} & {\tiny\textbf{\textcolor{red}{-48 \textpm 3 }}} & {\tiny\textbf{\textcolor{red}{-49 \textpm 3 }}} & {\tiny\textbf{\textcolor{red}{{} -30 \textpm 4 }}} & {\tiny\textbf{\textcolor{red}{{} -23 \textpm 3 }}} & {\tiny\textbf{\textcolor{red}{{} -33 \textpm 3 }}} & {\tiny{} -26 \textpm 2 } & {\tiny\textbf{\textcolor{red}{{} -15 \textpm 4 }}} & {\tiny\textbf{\textcolor{orange}{{} -77 \textpm 2}}}\tabularnewline
\hline 
\hline 
{\tiny$\mctsh{C=\frac{\sqrt{2}}{4}}$} & {\tiny\textbf{\textcolor{orange}{-55 \textpm 2}}} & {\tiny -66 \textpm 2} & {\tiny -50 \textpm 3} & {\tiny -26 \textpm 3} & {\tiny -67 \textpm 2} & {\tiny\textbf{\textcolor{red}{-8 \textpm 1}}} & {\tiny -48 \textpm 3} & {\tiny -48 \textpm 3}\tabularnewline
\hline 
{\tiny$\mctsh{C=\frac{\sqrt{2}}{8}}$} & {\tiny -56 \textpm 2} & {\tiny\textbf{\textcolor{orange}{-66 \textpm 2}}} & {\tiny\textbf{\textcolor{orange}{-46 \textpm 3}}} & {\tiny\textbf{\textcolor{orange}{-26 \textpm 3}}} & {\tiny\textbf{\textcolor{orange}{-62 \textpm 2}}} & {\tiny -9 \textpm 1} & {\tiny\textbf{\textcolor{orange}{-41 \textpm 3}}} & {\tiny\textbf{\textcolor{red}{-45 \textpm 3}}}\tabularnewline
\hline 
\hline 
{\tiny$\mcts{C=\frac{\sqrt{2}}{4}}$} & {\tiny -64 \textpm 2} & {\tiny -100 \textpm 0} & {\tiny -97 \textpm 0} & {\tiny -96 \textpm 0} & {\tiny -100 \textpm 0} & {\tiny -26 \textpm 1} & {\tiny -91 \textpm 1} & {\tiny -97 \textpm 0}\tabularnewline
\hline 
{\tiny$\mcts{C=\frac{\sqrt{2}}{8}}$} & {\tiny -64 \textpm 2} & {\tiny -99 \textpm 0} & {\tiny -97 \textpm 0} & {\tiny -96 \textpm 0} & {\tiny -100 \textpm 0} & {\tiny\textbf{\textcolor{orange}{-19 \textpm 1}}} & {\tiny -92 \textpm 1} & {\tiny -98 \textpm 0}\tabularnewline
\hline 
\end{tabular}}{\tiny\par}
\par\end{centering}
\caption{Binary scores for all games of the evaluated algorithms without batching
against $\protect\maxn$ with batching (search time $1s$ ; \textcolor{red}{red}
\textgreater{} \textcolor{orange}{orange} \textgreater{} \textcolor{pink}{pink}).}\label{tab:res_sans-batch-details-1}
\end{table*}
\begin{table*}
\begin{centering}
{\tiny{}%
\begin{tabular}{|c||c|c|c|c|c|c|c|c|}
\hline 
 & {\tiny Quadamazons} & {\tiny Blokus} & {\tiny Hey, That's My Fish!} & {\tiny Separed Teamhex} & {\tiny Quadrothello} & {\tiny 3 Player Hex} & {\tiny Triinversion} & {\tiny Threehex}\tabularnewline
\hline 
\hline 
{\tiny Safe Unbounded $\maxn$} & {\tiny\textbf{\textcolor{red}{-41 \textpm 3 }}} & {\tiny\textbf{\textcolor{orange}{{} -45 \textpm 3 }}} & {\tiny\textbf{\textcolor{red}{{} 0 \textpm 4   }}} & {\tiny\textbf{\textcolor{orange}{{} -18 \textpm 3 }}} & {\tiny\textbf{\textcolor{red}{{} -18 \textpm 3 }}} & {\tiny\textbf{\textcolor{red}{{} -2 \textpm 2  }}} & {\tiny\textbf{\textcolor{red}{{} -7 \textpm 4  }}} & {\tiny\textbf{\textcolor{red}{{} -25 \textpm 4}}}\tabularnewline
\hline 
{\tiny Unbounded $\maxn$} & {\tiny -47 \textpm 2 } & {\tiny{} -57 \textpm 2 } & {\tiny\textbf{\textcolor{orange}{{} -17 \textpm 3 }}} & {\tiny{} -34 \textpm 3 } & {\tiny\textbf{\textcolor{orange}{{} -37 \textpm 3 }}} & {\tiny{} -8 \textpm 1  } & {\tiny\textcolor{orange}{{} -29 \textpm 3 }} & {\tiny{} -40 \textpm 3}\tabularnewline
\hline 
{\tiny$\brs$} &  &  &  & {\tiny -20 \textpm 3 } & {\tiny{} -83 \textpm 1 } & {\tiny{} -5 \textpm 1  } & {\tiny{} -52 \textpm 3 } & {\tiny{} -31 \textpm 3}\tabularnewline
\hline 
{\tiny$\brsp$} & {\tiny\textbf{\textcolor{red}{-41 \textpm 2 }}} & {\tiny\textbf{\textcolor{orange}{{} -44 \textpm 2 }}} & {\tiny\textbf{\textcolor{pink}{{} -30 \textpm 3 }}} & {\tiny\textbf{\textcolor{orange}{{} -16 \textpm 3 }}} & {\tiny{} -78 \textpm 2 } & {\tiny\textbf{\textcolor{orange}{{} -5 \textpm 1  }}} & {\tiny{} -38 \textpm 3 } & {\tiny\textcolor{orange}{{} -34 \textpm 3}}\tabularnewline
\hline 
{\tiny Paranoid} & {\tiny\textbf{\textcolor{red}{-41 \textpm 2 }}} & {\tiny\textbf{\textcolor{red}{{} -42 \textpm 2 }}} & {\tiny{} -35 \textpm 3 } & {\tiny\textbf{\textcolor{orange}{{} -16 \textpm 3 }}} & {\tiny -79 \textpm 1 } & {\tiny\textbf{\textcolor{orange}{{} -5 \textpm 1  }}} & {\tiny{} -40 \textpm 3 } & {\tiny\textcolor{orange}{{} -33 \textpm 3}}\tabularnewline
\hline 
{\tiny$\maxn$} & {\tiny\textbf{\textcolor{orange}{-43 \textpm 2 }}} & {\tiny\textbf{\textcolor{pink}{{} -47 \textpm 2 }}} & {\tiny\textbf{\textcolor{pink}{{} -27 \textpm 3 }}} & {\tiny\textbf{\textcolor{orange}{{} -16 \textpm 3 }}} & {\tiny\textbf{\textcolor{pink}{{} -49 \textpm 2 }}} & {\tiny\textbf{\textcolor{orange}{{} -5 \textpm 1  }}} & {\tiny\textbf{\textcolor{pink}{{} -32 \textpm 3 }}} & {\tiny\textcolor{orange}{{} -34 \textpm 3}}\tabularnewline
\hline 
\hline 
{\tiny$k$-best $\maxn$ ($k=30$)} & {\tiny -45 \textpm 2 } & {\tiny{} -45 \textpm 2 } & {\tiny{} -25 \textpm 3 } & {\tiny{} -10 \textpm 3 } & {\tiny{} -49 \textpm 2 } & {\tiny{} -3 \textpm 1 } & {\tiny{} -35 \textpm 3 } & {\tiny{} -27 \textpm 3}\tabularnewline
\hline 
{\tiny$k$-best $\maxn$ ($k=16$)} & {\tiny -45 \textpm 2 } & {\tiny{} -48 \textpm 2 } & {\tiny{} -25 \textpm 3 } & {\tiny{} -12 \textpm 3 } & {\tiny{} -48 \textpm 2 } & {\tiny{} -3 \textpm 1  } & {\tiny{} -33 \textpm 3 } & {\tiny{} -26 \textpm 3}\tabularnewline
\hline 
{\tiny$k$-best $\maxn$ ($k=10$)} & {\tiny -42 \textpm 2 } & {\tiny{} -51 \textpm 2 } & {\tiny{} -16 \textpm 3 } & {\tiny{} -10 \textpm 3 } & {\tiny{} -49 \textpm 2 } & {\tiny{} -3 \textpm 1 } & {\tiny{} -33 \textpm 3 } & {\tiny\textbf{ -25 \textpm 3}}\tabularnewline
\hline 
{\tiny$k$-best $\maxn$ ($k=7$)} & {\tiny -46 \textpm 2 } & {\tiny{} -45 \textpm 2 } & {\tiny{} -16 \textpm 3 } & {\tiny{} -12 \textpm 3 } & {\tiny{} -51 \textpm 2 } & {\tiny{} -2 \textpm 1  } & {\tiny{} -31 \textpm 3 } & {\tiny{} -27 \textpm 3}\tabularnewline
\hline 
{\tiny$k$-best $\maxn$ ($k=4$)} & {\tiny -47 \textpm 2 } & {\tiny{} -49 \textpm 2 } & {\tiny{} -14 \textpm 3 } & {\tiny\textbf{\textcolor{red}{{} -9 \textpm 3  }}} & {\tiny{} -45 \textpm 2 } & {\tiny{} -4 \textpm 1  } & {\tiny{} -30 \textpm 3 } & {\tiny{} -28 \textpm 3}\tabularnewline
\hline 
{\tiny$k$-best $\maxn$ ($k=3$)} & {\tiny -46 \textpm 2 } & {\tiny{} -54 \textpm 2 } & {\tiny\textbf{ -10 \textpm 3 }} & {\tiny -12 \textpm 3 } & {\tiny\textbf{ -41 \textpm 2 }} & {\tiny{} -4 \textpm 1  } & {\tiny{} -28 \textpm 3 } & {\tiny{} -29 \textpm 3}\tabularnewline
\hline 
{\tiny$k$-best $\maxn$ ($k=2$)} & {\tiny -46 \textpm 2 } & {\tiny{} -65 \textpm 2 } & {\tiny{} -15 \textpm 3 } & {\tiny{} -15 \textpm 3 } & {\tiny{} -40 \textpm 2 } & {\tiny{} -3 \textpm 1  } & {\tiny{} -28 \textpm 3 } & {\tiny{} -27 \textpm 3}\tabularnewline
\hline 
\end{tabular}}{\tiny\par}
\par\end{centering}
\caption{Binary scores for all games of the evaluated algorithms with batching
against $\protect\maxn$ with batching (search time $1s$ ; \textcolor{red}{red}
\textgreater{} \textcolor{orange}{orange} \textgreater{} \textcolor{pink}{pink})}\label{tab:res_avec-batch-details-1}
\end{table*}
\begin{table}
\begin{centering}
{\small{}%
\begin{tabular}{|c||c|c|c|}
\hline 
 & {\small mean} & {\small lower bound} & {\small upper bound}\tabularnewline
\hline 
\hline 
{\small Safe Unbounded $\maxn$} & {\small\textbf{\textcolor{red}{-14.46}}} & {\small -15.69} & {\small -13.24}\tabularnewline
\hline 
{\small Unbounded $\maxn$} & {\small\textbf{\textcolor{pink}{-35.54}}} & {\small -36.45} & {\small -34.63}\tabularnewline
\hline 
{\small$\brsp$} & {\small -37.41} & {\small -38.29} & {\small -36.54}\tabularnewline
\hline 
{\small Paranoid} & {\small -38.02} & {\small -38.91} & {\small -37.15}\tabularnewline
\hline 
{\small$\maxn$} & {\small\textbf{\textcolor{orange}{-33.05}}} & {\small -33.98} & {\small -32.14}\tabularnewline
\hline 
\hline 
{\small$k$-best $\maxn$ (best $k=4$)} & {\small -29.77} & {\small -30.71} & {\small -28.85}\tabularnewline
\hline 
\end{tabular}}{\small\par}
\par\end{centering}
\caption{Average binary scores over all games of the evaluated algorithms (except
$\protect\brs$) with batching against $\protect\maxn$ with batching
(search time $1s$ ; \textcolor{red}{red} \textgreater{} \textcolor{orange}{orange}
\textgreater{} \textcolor{pink}{pink}).}\label{tab:res_avec-batch-mean-1}
\end{table}

\subsection{Games}\label{subsec:Games}

We now detail the rules of the games used as benchmarks for our experiments.
To avoid any ambiguity, the games we introduce are also provided with
their source code in supplementary materials (otherwise we provide
a link to the official rules).

\subsubsection{Blokus}

Blokus is a classic abstract game. It is a game of tile placement,
enclosure, chaining, and hand management. More precisely, it is an
abstract strategy game played on a square grid board. Each player
is assigned a distinct color and a fixed set of 21 polyomino pieces
of varying shapes and sizes. Players alternately place one piece of
their color onto the board, subject to strict spatial constraints.

A legal placement requires that the newly placed piece does not share
an edge with any previously placed piece of the same color; instead,
it must touch at least one corner of an existing piece of that color.
Pieces of different colors may touch freely along edges or corners.

The objective is to maximize the number of pieces placed on the board.
The game ends when no player can make a legal move, and the score
is determined by the total number of pieces successfully placed.

The standard game is played on a 20\texttimes 20 grid (400 squares)
and includes 84 polyomino pieces, divided into four color sets of
21 pieces each. For each color, the set consists of one monomino,
one domino, two trominoes, five tetrominoes, and twelve pentominoes.

It is ranked as the 61th best abstract game (on the boardgamegeek
website ranking). 

Blokus has been used on multiple occasions as a benchmark for artificial
intelligence algorithms \cite{chao2018blokus,nijssen2013monte,nijssen2013search,hartjes2019feasibility,schadd2011best}.

Complete rules of Blokus are available \href{https://boardgamegeek.com/boardgame/2453/blokus/files}{here}.

Note that the associated terminal evaluation that we used is the score
of the game.

\subsubsection{Hey, That's My Fish!}

Hey, That’s My Fish! is a game of Grid Movement, Map Reduction with
Variable Set-up. More precisely, it is an abstract strategy game in
which players control multiple penguin tokens moving on a hexagonal
grid of ice tiles. Each tile is labeled with a fixed number of fish
(one to three), which determine the scoring potential of that tile.
The tiles are initially placed at random.

On each turn, a player selects one of their penguins and moves it
in a straight line across any number of contiguous ice tiles, following
hexagonal directions. After the move, the tile from which the penguin
departed is removed from the board, creating an impassable gap that
permanently alters the game topology. In addition, the player collects
the removed tile (on which their token was located). When a penguin
has no legal moves remaining, it is removed from play.

The game ends when all penguins have been removed. The winner is the
player who has accumulated the largest total number of fish across
all collected tiles.

It is ranked as the 81th best abstract game and the 391 best family
game (on the boardgamegeek website ranking). 

Complete rules of Hey, That’s My Fish! are available \href{https://images-cdn.fantasyflightgames.com/ffg_content/hey-thats-my-fish-board-game/hey-thats-my-fish-rulebook.pdf}{here}

Note that the associated terminal evaluation that we used is the score
of the game.

\subsubsection{3-Player Hex}

Three-Player Hex is played on a board with hexagonal cells and overall
hexagonal shape. In this study, boards have 7 cells per side. 

Players alternate turns by placing a stone of their assigned color
on an unoccupied cell. The objective, as in standard Hex, is to form
a continuous chain of stones connecting the two opposite sides of
the board designated for their color.

There are two variations of this game. In the first, which we have
implemented, the game ends in a draw when none of the three players
can connect their edges. In the other variation, as soon as a player
cannot connect their edges, they are eliminated, and as soon as two
players are eliminated, the third wins.

Complete rules of 3-Player Hex are available on \href{https://ludii.games/details.php?keyword=Three-Player\%20Hex}{Ludii}.

Note that the presence of a third player eliminates the connection
guarantees of strategics patterns that are the basis of two-player
hex strategies

Note that the associated terminal evaluation that we used is the additive
deep heuristic~\cite{cohen2020learning}.

\subsubsection{Threehex }

We introduce an alternative three-player generalization of Hex designed
to better preserve the strategic properties of the two-player game,
by significantly reducing the joint interaction between the two opponents
that arises in the classical three-player variant of Hex.

The objective of the game remains unchanged: a player wins by being
the first to connect their two designated sides of the board with
a chain of pieces of their color. However, it is now allowed to place
a piece on a cell that already contains a piece belonging to another
player, except when that player is the next player in turn.

More precisely, player 0 may place a piece on an empty cell or on
a cell containing a single piece belonging to player 2; player 1 may
place a piece on an empty cell or on a cell containing a single piece
belonging to player 0; and player 2 may place a piece on an empty
cell or on a cell containing a single piece belonging to player 1.
A player who has no legal move must pass. If all players pass consecutively,
the game ends in a draw.

Note: if a piece is overlaid by another piece, it still counts toward
connecting the corresponding sides of the board. The length of each
player’s designated board edges is 7 in these experiments.

With this modification, only one of the two opponents can prevent
a given player from connecting their two sides, namely the next player
in turn. As a result, each player primarily aims to block their predecessor
while simultaneously attempting to complete their own connection.
Symmetrically, each player is themselves blocked by their successor.
This interaction structure partially restores strategic patterns characteristic
of the two-player game.

In addition, indirect interactions arise when attempting to counter
the player who is blocking us. Depending on where a player places
pieces to block their predecessor, that predecessor may respond by
placing pieces in locations that, in turn, block the third player.
Consequently, players must act so as to block their immediate predecessor
as effectively as possible, while anticipating that the opponent’s
response will hinder the remaining player. This leads to bidirectional
blocking interactions among all three players.

\subsubsection{Separed Teamhex}

We introduce the game Separed Teamhex as a 4-player generalization
of the game Hex.

\paragraph{Game Overview}

Separed Teamhex is a connection game played by four players on a square
hexagonal board (same board as in two-player hex).

Players are arranged into two fixed teams of two, and victory can
be achieved either individually or jointly with an ally.

\paragraph{Board and Pieces}

\begin{figure}
\begin{centering}
\includegraphics[scale=0.12]{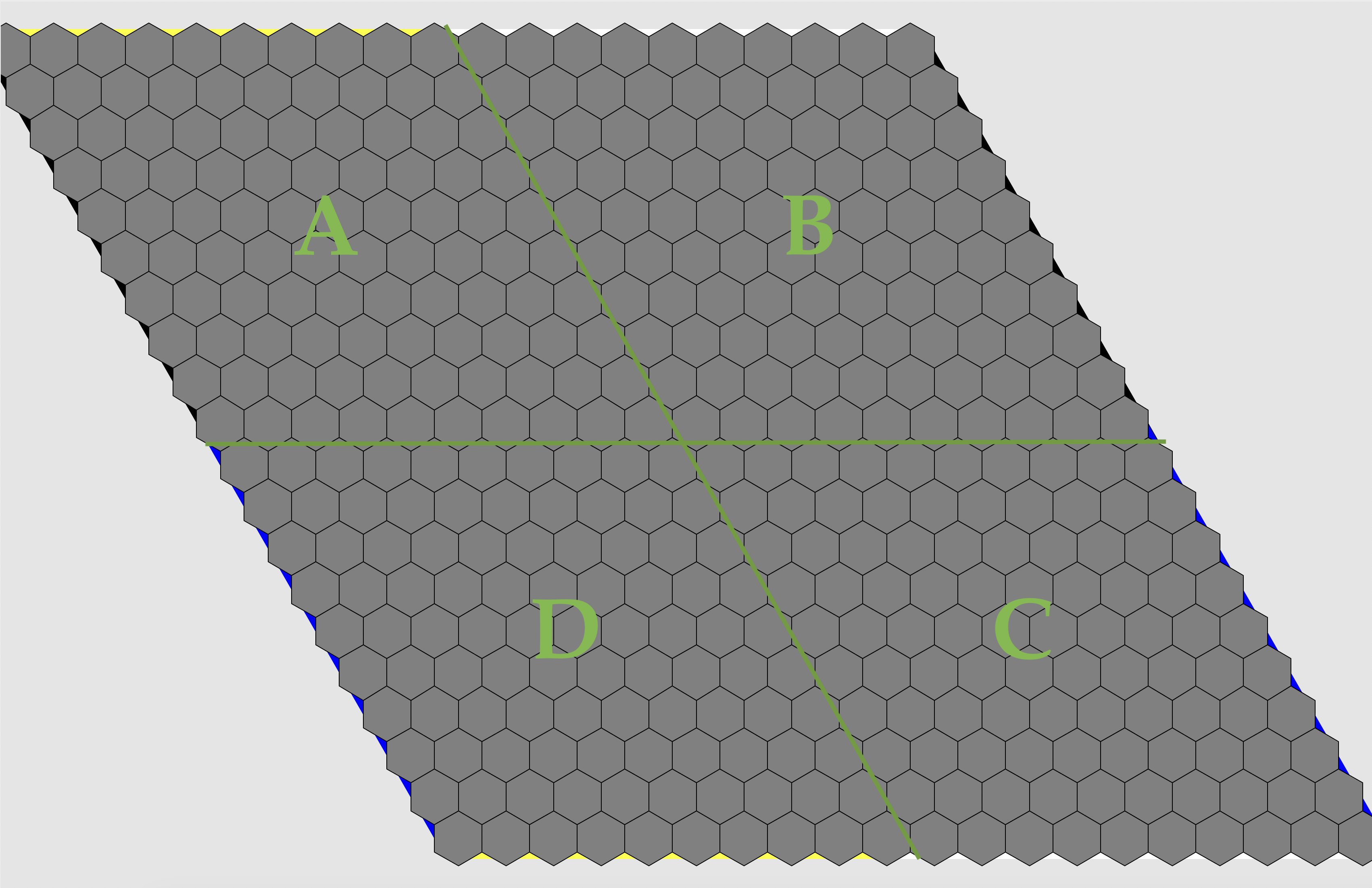}
\par\end{centering}
\caption{Board for Separed Teamhex}\label{fig:quadhex-separer}
\end{figure}

The game is played on an $N\times N$ square hexagonal board, with
$N$ even (default $N=20$). Each cell can contain at most one stone.
Each player has a distinct stone color. The board is divided into
zones A, B, C, and D (see Figure~\ref{fig:quadhex-separer}).

\paragraph{Players and Teams}

There are four players, indexed $\{0,1,2,3\}$. Teams are fixed:
\begin{itemize}
\item Team A: players 0 and 2
\item Team B: players 1 and 3
\end{itemize}
Stones of a player and their ally are considered connected for the
purpose of path formation for shared wins.

\paragraph{Turn Order, Color, and Moves}

Player $0$ is Back, player 1 is White, player 2 is Blue, player 3
is Yellow.

In turn, a player can only play a piece of their color on any empty
cell in a specific zone. In each zone, only two players can play (the
rules of two-player Hex are found locally). The turn order and the
allowed zones associated with the corresponding action are as follows:
black player plays in zone A, then yellow player plays in zone A,
then white player plays in zone B, then black plays in zone B, then
blue plays in zone C, then white plays in zone C, then yellow plays
in zone D, then blue plays in zone D. This cycle repeats until the
end of the game. The game ends immediately when a winning condition
is met.

\paragraph{Adjacency and Connectivity}

Connectivity follows Hex-style adjacency: each cell has up to six
neighbors.

\paragraph{Player-Specific Goal Edges}

Each player is assigned two disjoint segments of the outer board boundary,
i.e. of the board sides (see Figure~\ref{fig:quadhex-separer}).
Team sides are the union of team members sides (we find the double
pair of opposite edges from the two-player game). 

\paragraph{Winning Conditions}

After each move, the game checks whether a connected component (using
Hex adjacency) touches two opposite board sides for at least one player
or for at least one team. Two types of victory are distinguished.

Strong (Individual) Win: a player achieves a strong win if their own
stones alone form a connected path between their two sides.

Team (Shared) Win: if no strong win exists, but a connected path exists
between the two sides of the team using stones from both allied players,
then both allies win jointly.

\paragraph{Scoring}
\begin{itemize}
\item Strong individual win:
\begin{itemize}
\item Winning player: +2
\item All other players: \textminus 2
\end{itemize}
\item Team win:
\begin{itemize}
\item Each winning ally: +1
\item Each opposing player: \textminus 1
\end{itemize}
\end{itemize}
Note that the associated terminal evaluation that we used is the score
of the game.

\paragraph{Termination}
\begin{itemize}
\item The game always terminates immediately upon the first detected winning
configuration.
\item There are no draws.
\end{itemize}

\subsubsection{Quadamazons}

We introduce the following generalization of the game Amazons for
4 players.

\paragraph{Board}

The board is a square grid of size $N\times N$ ($N$ even ; default
$N=14$) with a multi-interaction distance parameter $d$ (default
$d=2$). The parameter $d$ divides the board into 9 atomic zones:
4 zones of size $\left(\frac{N}{2}-d\right)\times\left(\frac{N}{2}-d\right)$
where only two players can play; 2 zones $\left(\frac{N}{2}-d\right)\times\left(2\cdot d\right)$
and 2 zones $\left(2\cdot d\right)\times\left(\frac{N}{2}-d\right)$
where three players can play and one zone $\left(2\cdot d\right)\times\left(2\cdot d\right)$
where all 4 players can play. As in the classic game, cells can be
empty, occupied by a player piece called amazon or blocked by an arrow
(a neutral permanent obstacle piece).

\paragraph{Initial Setup}

\begin{figure}
\begin{centering}
\includegraphics[scale=0.2]{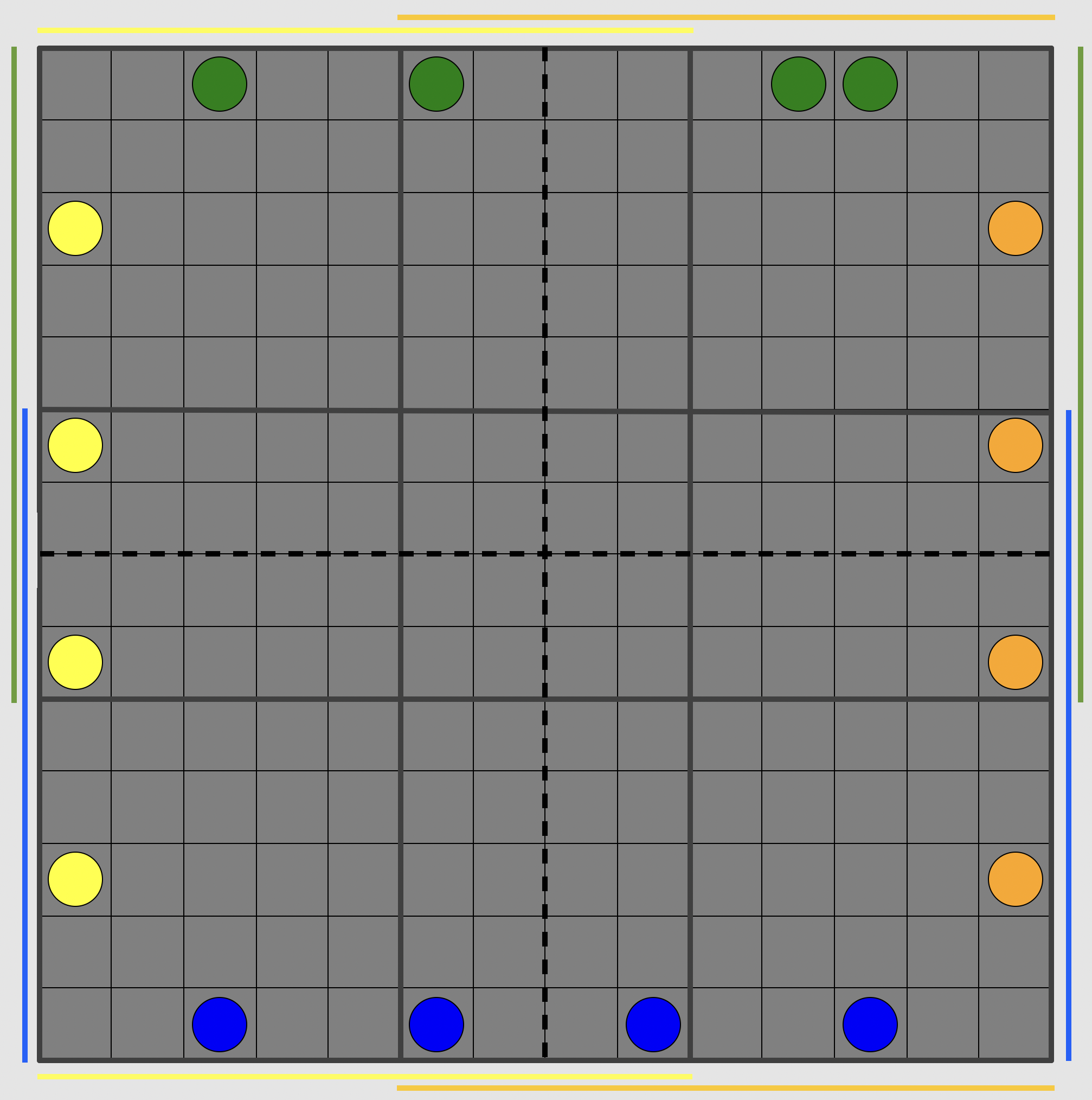}
\par\end{centering}
\caption{}\label{fig:quadamazons}

\end{figure}

Each player starts with 4 amazons pieces. Initial positions are fixed
on the board edges (see Figure~\ref{fig:quadamazons}).

\paragraph{Actions}

The game proceeds in turns. Each turn is composed of two steps and
each step is composed of two phases.

The first phase of each step consists in moving one of their amazon:
all amazons move like a chess queen (possible movements: in a straight
line in one of the 8 directions, at any distance, to arrive at an
empty square, but cannot pass over an obstacle: i.e. an amazon or
an arrow). Each amazon can only move within the area associated with
their player: the rectangle $\left(\frac{N}{2}+d\right)\times N$
or $N\times\left(\frac{N}{2}+d\right)$ corresponding to the space
on the board surrounded by the two lines of the player's color (see
Figure~\ref{fig:quadamazons}). Each player zone is divided into
two sub-zones by a dotted line (two sub-zones of size $\left(\frac{N}{2}+d\right)\times\frac{N}{2}$
or $\frac{N}{2}\times\left(\frac{N}{2}+d\right)$). If the first amazon
moved during the current turn arrives in one of the two sub-zones,
the second amazon to move must end up in the other sub-zone.

The second phase of each step consists in placing an arrow as if it
had been pulled by the amazon that had just moved (as in the classic
game). The legal actions for an arrow are any empty case starting
from the new position of the amazon by following the legal movement
of the chess queen. In addition, the arrow must be placed in the sub-zone
where the amazon is newly located or be placed in the exact position
where the amazon was before its movement.

\paragraph{Elimination and Turn Order}

If a player has no legal move at the first phase of their first step,
it is eliminated. Eliminated players are skipped. If a player has
no legal move at the first phase of their second step, their step
is passed. The game continues until all players are eliminated, or
only one player remains with legal moves.

\paragraph{End of Game and Scoring}

The game ends in two cases:
\begin{enumerate}
\item All players are eliminated: the last player to be eliminated is declared
the winner. Scores are assigned by reverse elimination order: last
eliminated player scores $1$, the second last eliminated player scores
$0$, third last player scores $-1$, and the first eliminated player
scores $-2$.
\item One player remains: the remaining player is the winner. Their score
equals the number of legal moves available at termination: $s$. The
last eliminated player scores $0$, the second last eliminated player
scores $-s$, and the first eliminated player scores $-2s$.
\end{enumerate}
Note that the associated terminal evaluation that we used is the score
of the game.

\subsubsection{Quadrothello}

We introduce the following generalization of Othello (also known as
Reversi) for four players, that we called Quadrothello. A multiplayer
version of Othello, called Rolit, already exists. However, this game
suffers from two major flaws: unlike the two-player game, an eliminated
player can return to the game, and the game is extremely chaotic,
making planning impossible in practice. The generalization we propose
allows us to significantly increase control in order to be a more
interesting benchmark for comparing search algorithms.

\paragraph{Overview}

As in the two-player game, players compete to control as many board
cells as possible when the game ends. However, players are restricted
to specific zones, meaning that only two players can play in any given
zone. This allows for a localized version of the two-player Othello
rules, thus preserving its control and strategies. However, actions
performed in one zone can affect other zones, making it a true multiplayer
game.

\paragraph{Board and Players}

The game is played on a square board of size $N\times N$ ($N$ even,
default $N=14$).

\begin{figure}
\centering{}\includegraphics[scale=0.2]{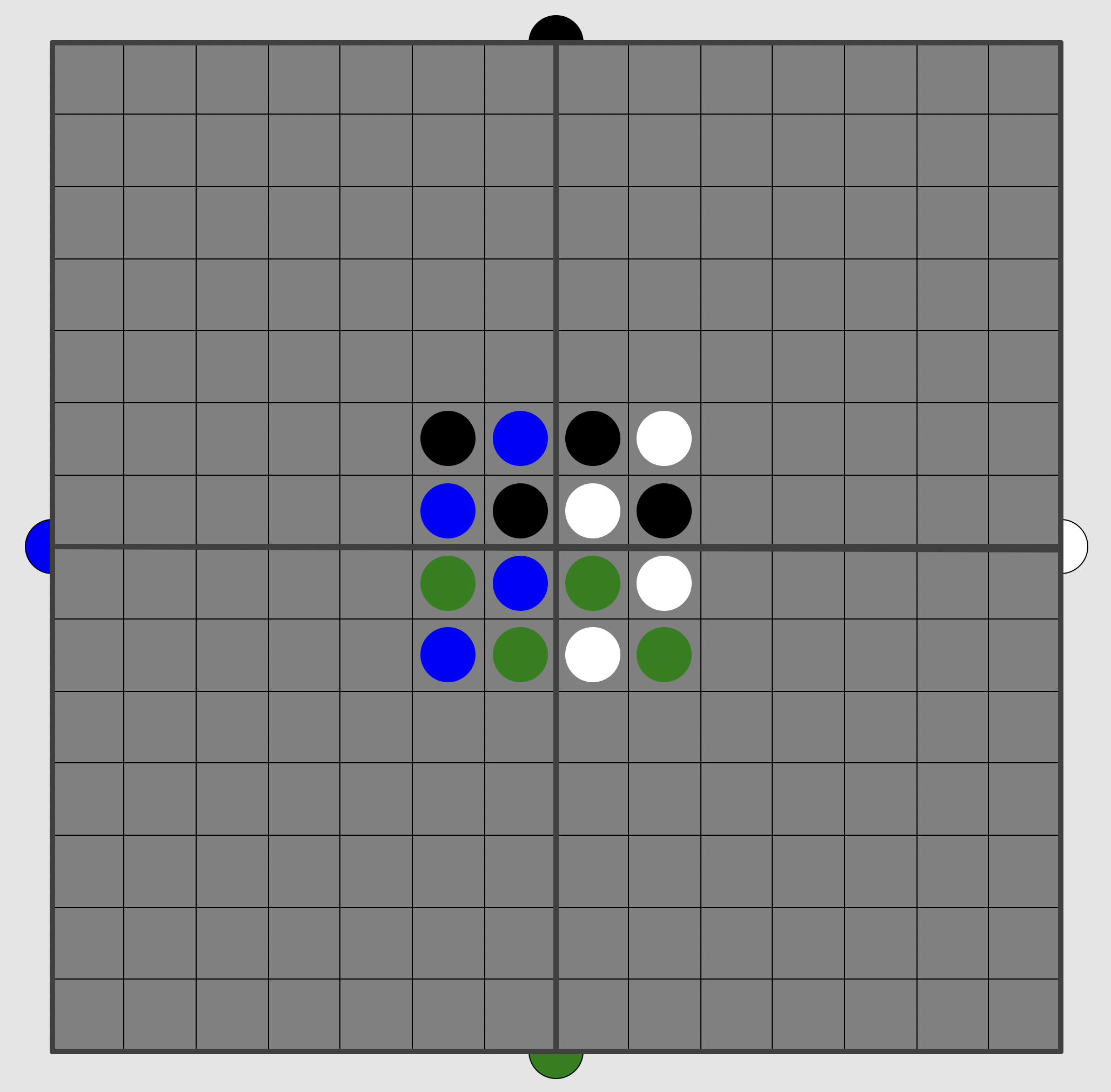}\caption{Quadrothello board at the start of the game (the semicircles indicate
the players' zones).}\label{fig:quadrothello}
\end{figure}

The board is partitioned into four overlapping zones of shape $N\times\frac{N}{2}$
or $\frac{N}{2}\times N$, one per player (see Figure~\ref{fig:quadrothello}):
\begin{itemize}
\item Player 0 (black): top half
\item Player 1 (white): right half
\item Player 2 (green): bottom half
\item Player 3 (blue): left half.
\end{itemize}

Each player starts with four stones, placed near the center of the
board in a local symmetric configuration.

\paragraph{Turn Order}

Players take turns in cyclic order: $0\rightarrow1\rightarrow2\rightarrow3\rightarrow0\rightarrow\dots$.
If a player has no legal moves, their turn is automatically skipped.

\paragraph{Legal Moves}

A move consists of placing a stone on an empty cell inside the current
player’s zone. A move is legal if and only if the cell is empty and
there exists at least one direction (among the 8 surrounding directions:
horizontal, vertical, diagonal) such that:
\begin{itemize}
\item One or more consecutive opponent stones are encountered,
\item Followed by a stone belonging to the current player.
\end{itemize}
After placing a stone, all opponent stones between the two player
stones are captured (remplaced by current player stones).

\paragraph{Game End}

The game ends when all four players consecutively have no legal moves.

\paragraph{Scoring and Winner}

Each player’s score is the number of stones they own on the board.
At game end, the player(s) with the maximum score win. Ties are allowed
(multiple winners).

Note that the associated terminal evaluation that we used is the score
of the game.

\subsubsection{Triinversion}

We propose the following other generalization of Othello, this time
for three player, that we called Triinversion. This is no longer exactly
a generalization but a variation: the game is played on a hexagonal
board, so there are only 6 directions for alignment. Unlike Quadrothello,
there are no spatial constraints on placement. However, in order to
have more control over the captures of pieces, the capture rules have
been modified.

\paragraph{Board and Players}

The game is played by three players, indexed as Player 0 (cyan), Player
1 (magenta), and Player 2 (yellow). The board is a hexagonal hex board
of side length $l$ (default parameter length $l=6$). Each cell can
contain at most one piece, owned by one of the three players.

Special adjacency rules apply at the geometric center to preserve
hexagonal connectivity: the central position is not a playable position
and all pair of positions adjacent to the center, which which are
opposite each other with respect to the center, are considered adjacent
(and therefore aligned for captures).

Each player has a direct opponent and an indirect opponent. Player
0's direct opponent is player 2, player 1's direct opponent is player
0, and player 2's direct opponent is player 1. The other opponent
is the indirect opponent.

\paragraph{Initial Setup}

\begin{figure}

\begin{centering}
\includegraphics[scale=0.2]{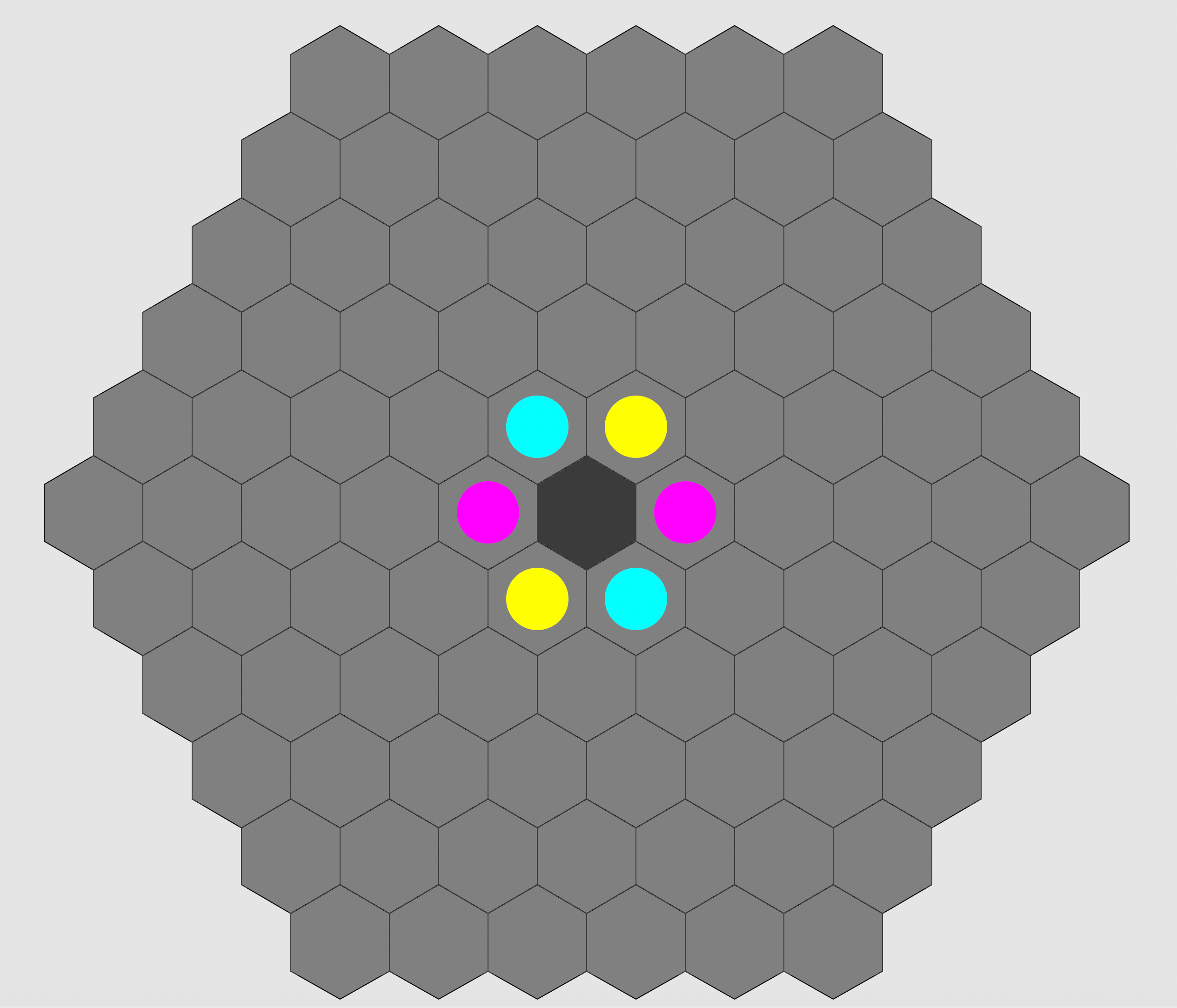}\caption{Triinversion board at the start of the game.}\label{fig:triinversion}
\par\end{centering}
\end{figure}

At the beginning of the game, six pieces are placed symmetrically
around the center: see Figure~\ref{fig:triinversion}.

\paragraph{Turn Order}

Players take turns in cyclic order: Player 0 → Player 1 → Player 2
→ Player 0 → \ldots . If a player has no legal moves, they must pass.
If all three players consecutively have no legal moves, the game ends.

\paragraph{Legal Moves}

A legal action is to place a piece of their color on an empty cell
in such a way that a line of their direct opponent's pieces is surrounded
by that piece and either one of their pieces or one of their indirect
opponent's pieces. By placing such a piece, all of their direct opponent's
pieces are removed and replaced by their pieces. 

As a reminder, two alignments of pieces of the same player in the
same direction separated only by the always empty central cell are
considered as a single alignment and can therefore be surrounded and
replaced all at once.

\paragraph{Game Termination}

The game ends when all three players consecutively have no legal moves.

\paragraph{Winner Determination and Scoring}

The winner is the player with the highest final score. In case of
a tie, the tied players are jointly considered winners. A player's
score is their number of pieces plus the number of pieces of their
indirect opponent.

Note that the associated terminal evaluation that we used is the score
of the game.
\end{document}